\documentclass{elsart}
\usepackage[dvips]{graphicx}

\begin{document}

\begin{frontmatter}
\title{A new form of governing equations of fluids \\
arising from Hamilton's principle}
\author[Marseille]{S. Gavrilyuk} and
\author[Marseille]{H. Gouin}

\address[Marseille]{Laboratoire de Mod\'elisation en M\'ecanique et Thermodynamique,
Facult\'e des Sciences, Universit\'e d'Aix - Marseille III, Case
322, Avenue Escadrille Normandie-Niemen, 13397 Marseille Cedex 20,
FRANCE}

\begin{abstract}
A new form of governing equations is derived from  Hamilton's
principle of least action for a constrained Lagrangian, depending on
conserved quantities and their derivatives with respect to the
time-space. This form yields conservation laws both for
non-dispersive case (Lagrangian depends only on conserved
quantities) and dispersive case (Lagrangian depends also on their
derivatives).
 For non-dispersive case the set of conservation laws allows to rewrite the governing equations
in the
 symmetric form of
Godunov-Friedrichs-Lax.
 The linear stability of equilibrium states for potential motions is also studied. In particular,
 the dispersion relation is obtained in terms of
Hermitian matrices both for non-dispersive and
 dispersive case.
Some new results are extended to the two-fluid non-dispersive case.
\end{abstract}

\end{frontmatter}

\section{Introduction}

Hamilton's principle of least action is frequently used in conservative fluid mechanics
[1-3]. Usually, a given Lagrangian
\ $\Lambda$ \  is submitted to constraints representing conservation in the time-space of
collinear vectors
\ ${\bf j}_k$
$$
{\rm Div} \, {\bf j}_k \  = \ 0, \quad k=0,{\dots},m
\eqno{(1.1)}
$$
where \ ${\rm Div}$ \ is the divergence operator in the time-space. Equation (1.1) means the
conservation of mass, entropy, concentration, etc.

Lagrangian appears as a function of \ ${\bf j}_k$ \ and their derivatives.
To calculate the variation of Hamilton's action we don't use
Lagrange multipliers to take into account the constraints (1.1).
 We use the same method as Serrin in [2] where the variation of the density
\ $\rho$ \ is expressed directly in terms of the virtual displacement of the medium.
This approach yields an antisymmetric form for the governing equations
$$
\displaystyle
\sum^m_{k=0} \ {\bf j}_k^{*}
\left( {\partial {\bf K}_k \over \partial {\bf z}}  \ - \
\left( {\partial {\bf K}_k \over \partial {\bf z}} \right)^{*} \right) \ = \ 0
\eqno{(1.2)}
$$
where \ ${\bf z}$ \ is the time-space variable, "star" means the transposition, and
\ ${\bf K}_k$ \ is the variational derivative
of \ $\Lambda$ \ with respect to \ ${\bf j}_k$
$$
{\bf K}_k^{*} \ =\ {\delta \Lambda \over \delta {\bf j}_k}
\eqno{(1.3)}
$$
Equations (1.1) - (1.3) admit particular class of solutions called potential flows
$$
{\bf j}_k \ = \ a_k {\bf j}_0,
\quad  a_k \ = \ {\rm const}, \quad k=1,{\dots},m
$$
$$
{\bf K}_0^{*}  \ = \ {\partial \varphi_{0} \over \partial {\bf z}}
$$
where \ $\varphi_{0}$\
is a scalar function. We shall study the linear stability of constant solutions for potential flows.

In Section 2, we present the variations of unknown quantities in terms of virtual displacements
of the continuum.
In Section 3, we obtain the governing system (1.2) using Hamilton's principle of least action.
Conservation laws admitted by the system (1.1) - (1.3) are obtained in section 4.
For non-dispersive case,
 we obtain the equations (1.1) - (1.3)
in the symmetric form of Godunov-Friedrichs-Lax [4,5].
Section 5 is devoted to the linear stability of equilibrium states (constant solutions)
for potential motions.
 We obtain dispersion relations in terms of Hermitian matrices both for dispersive
and non-dispersive flows and
propose simple criteria of stability. In Section 6-7, we generalize our results
for two-fluid mixtures in the
non-dispersive case. Usually, the theory of mixtures considers two different cases of
continuum media. In {\it homogeneous} mixtures such as binary gas mixtures, each component
occupies the whole volume of a physical space. In {\it heterogeneous} mixtures such as a
mixture of incompressible liquid containing gas bubbles, each component occupies a
part of the volume of a physical space. We do not distinguish the two cases because they
have the same form for the governing equations. We obtain a simple stability
criterion (criterion of hyperbolicity) for small relative velocity of phases.
In Appendix, we prove non-straightforward calculations.

Recall that "star" denotes {\it conjugate} (or {\it transpose}) mapping or covectors (line vectors).
For any vectors \ ${\bf a, b}$ \
we shall use the notation \ ${\bf a^{*} b}$ \ for their {\it scalar product}
(the line vector is multiplied by the column vector) and
\ ${\bf a b^{*}}$ \ for their
{\it tensor product} (the column vector is multiplied by the line vector). The product of a mapping
\ $A$\  by a vector \ ${\bf a}$ \ is denoted by \ $A {\bf a}$,\, ${\bf b}^{*} A$\  means covector
\ ${\bf c}^{*}$ \ defined by the rule
\ ${\bf c}^{*} = (A^{*}{\bf b})^{*}$. The divergence of a linear transformation \ $A$\  is
the covector \ ${\rm Div} \, A$ such that, for any constant vector \ ${\bf a}$,
$$
{\rm Div}\, (A) {\bf a} \  = \ {\rm Div}\, (A {\bf a})
$$
The identical transformation is denoted by \ $I$.
 For divergence and gradient operators in the time-space we use respectively
symbols \ ${\rm Div}$ \ and
$\displaystyle \ {\partial \over \partial {\bf z}}$, where $\displaystyle \ {\bf z}^{*}
\ =\ (t, {\bf x}^{*})$, \ $t$ \
is the time and \ ${\bf x}$\  is the space.
The gradient line (column) operator in the space is denoted by \ $\nabla\, ({\nabla}^{*})$, and
the divergence operator in the space
 by \ ${\rm div}$.
The elements of the matrix \ $A$\  are denoted by \ $a^i_j$\  where \ $i$\  means lines and
\ $j$\
 columns. If \ $f(A)$\  is a scalar function of
\ $A$,\  matrix \ $\displaystyle\ B^{*} \equiv \ {\partial f \over \partial A}$\
is defined by the formula
$$
\left( B^{*} \right)^j_i \ = \ \left( {\partial f \over \partial A }\right)^j_i \
=\ {\partial f\over \partial a^i_j}
$$
  The repeated latin indices
mean summation. Index \ $\alpha\, =\,1,2$\  refers to the parameters of components
densities \ $\rho_\alpha$,  velocities \ ${\bf u}_\alpha$, etc.

\section{Variations of a continuum}

Let \ ${\bf z}\ = \ \displaystyle \bigg(\matrix{t\cr
{\bf x}}\bigg)$\  be Eulerian coordinates of a particle of a continuum and
 \ $D(t)$\  a volume of the physical space occupied by a fluid at time \ $t$.  When \ $t$\
belongs to a finite interval \ $[t_0\,, \,t_1]$, \,  $D(t)$\  generates a
four-dimensional domain \ $\Omega$\  in the time-space.
A particle is labelled by its position \ ${\bf X}$\  in a reference space \ ${ D}_0$.  For example,
if \ $D(t)$ \ contains always the same particles \ $D_0 \,  =\, D(t_0)$,  and we can define the
motion of a continuum as a diffeomorphism from \ $D(t_0)$\  into \ $D(t)$\
$$
{\bf x} \  =\ {\mbox{{\boldmath $\chi$}}}_t \, ({\bf X})
\eqno{(2.1)}
$$
We generalize (2.1) by defining the motion
as the diffeomorsphism
from the reference space \ $\Omega_0 $\  into the time-space \ $\Omega $\ occupied by the medium
in the following parametric form
$$
\displaystyle \cases{ t \  =\ g\, (\lambda \,, \,{\bf X})\cr\cr
 {\bf x} \  =\ {\mbox{{\boldmath $\phi$}}} \, (\lambda \,, \,{\bf X})}
\eqno{(2.2)}
$$
where \ ${\bf Z}\, = \displaystyle \bigg(\matrix{\lambda \cr
{\bf X}}\bigg) $\  belongs to a reference space
 \ $\Omega_0$.
The mappings
$$
\displaystyle \cases{\lambda \  =\ h \,(t \,, \,{\bf x})\cr\cr
{\bf X} \  =\ { \mbox{{\boldmath $\psi$}}} (t \,, \,{\bf x})}
\eqno{(2.3)}
$$
are the inverse of (2.2).
Definitions (2.2) imply the following expressions for the differentials
\ $dt$\  and \ $d{\bf x}$\
$$
\bigg(\matrix{dt\cr
d{\bf x}}\bigg) \  =\  B \, \bigg(\matrix{d\lambda\cr
d{\bf X}}\bigg)
\eqno{(2.4)}
$$
where
$$
B \  =\  \left( \matrix{\displaystyle \ {\partial g\over \partial
\lambda}\,\,,\,\,{\partial g\over \partial{\bf X}}
\cr\cr
\displaystyle {\partial {\mbox{{\boldmath $\phi$}}} \over \partial \lambda}\,\,,\,\,
{\partial {\mbox{{\boldmath $\phi$}}}\over \partial {\bf X}} } \right)
$$
Formulae (2.2), (2.4) assume the
form
$$
 \displaystyle \cases{\displaystyle dt \  = \  { \partial g \over \displaystyle
\partial \lambda}\,\,d\lambda \ +\ {\partial g\over
\partial {\bf X}}\ d{\bf X}\cr\cr
\displaystyle d{\bf x} \  =\ {\partial { \mbox{{\boldmath $\phi$}}}\over \partial \lambda}\ d\lambda \ +
\ {\partial { \mbox{{\boldmath $\phi$}}}\over \partial {\bf X}}\ d{\bf X}}
\eqno{(2.5)}
$$
From equation (2.5) we obtain
$$
d {\bf x} \ =\ {\bf u}\, dt \ +\ F\, d {\bf x}
$$
where velocity \ ${\bf u}$\  and deformation gradient \ $F$\  are defined by
$$
{\bf u} \ =\ {\partial { \mbox{{\boldmath $\phi$}} }\over \partial \lambda} \,
\left( {\partial g\over \partial
\lambda} \right)^{-1} \,, \quad F \ =\ {\partial { \mbox{{\boldmath $\phi$}} }\over \partial {\bf X} }
\ -\ {\partial { \mbox{{\boldmath $\phi$}} }\over
\partial \lambda}\, {\partial g\over \partial {\bf X}}\, \left( {\partial g\over \partial
\lambda} \right)^{-1}
\eqno{(2.6)}
$$
Let
$$
 \displaystyle \cases{t\  =\ G\, (\lambda \,,  \,{\bf X}  \,,  \,\varepsilon)\cr\cr
 {\bf x}\  =\ {\bf \Phi}\, (\lambda \,,  \,{\bf X}  \,, \,\varepsilon)}
\eqno{(2.7)}
$$
be a one-parameter family of virtual motions of the medium such that
$$
G\, (\lambda  \,, \,{\bf X}  \,, \,0)\  =\ g(\lambda  \,,  \,{\bf X}), \quad
{\bf \Phi}\, (\lambda  \,,  \,{\bf X}  \,, \,0)\  =\ { \mbox{{\boldmath $\phi$}}}\, (\lambda \,,
\, {\bf X})
$$
 where \ $\varepsilon$\  is a scalar defined in the vicinity of zero. We define Eulerian
displacement \ ${ \mbox{{\boldmath $\zeta$}}}\  =\ ( \tau  \,,  \,{ \mbox{{\boldmath $\xi$}}})$\
associated with the virtual motion (2.7)
$$
\tau\  =\ {\partial G\over \partial \varepsilon}\, (\lambda  \,,  \,{\bf X}  \,, \,0),
 \quad  {\mbox{{\boldmath $\xi$}}}\  =\ {\partial {\bf \Phi}\over \partial \varepsilon}\, (\lambda
 \,,  \, {\bf X}  \,,  \,0)
\eqno{(2.8)}
$$
We note that \ ${ \mbox{{\boldmath $\zeta$}}} $ is naturally defined in Lagrangian coordinates. However, we
shall suppose that \ ${ \mbox{{\boldmath $\zeta$}}} $ is represented in Eulerian coordinates by
means of (2.3).

Let us now consider any tensor quantity represented by \ $f\,(t, \,{\bf x})$\
in Eulerian
coordinates and \ $\buildrel\circ\over {f}\, (\lambda \,, \,{\bf X})$\  in
Lagrangian coordinates. Definitions (2.2), (2.3) involve
$$
\buildrel\circ\over {f}\,(\lambda \,, \,{\bf X}) \  =\ f\, \bigg( g\,(\lambda,
\,{\bf X})\,, \,{ \mbox{{\boldmath $\phi$}}} \,(\lambda, \,{\bf X}) \bigg)
\eqno{(2.9)}
$$
Conversely,
$$
f\, (t\,, \,{\bf x}) \  =\ {\buildrel\circ\over {f}}\, \bigg( h\, (t, \,{\bf x})
\,, \,{ \mbox{{\boldmath $\psi$}}} \,(t, \, {\bf x}) \bigg)
\eqno{(2.10)}
$$

Let \ $\buildrel\sim\over{f}\, (\lambda ,  \,{\bf X} ,  \,\varepsilon)$\  and \
$\buildrel\wedge\over{f} \,(t , \, {\bf x}  ,  \,\varepsilon)$\  be tensor quantities associated
with the  virtual motions, such that \ $\buildrel\sim\over{f}\, (\lambda,
\,{\bf X},
\,\varepsilon) \ \equiv \ \buildrel\wedge\over{f} \,(t, \, {\bf x}, \,\varepsilon)$\
 where \ $\lambda$,\, ${\bf X}$,\, $t$,\, ${\bf x}$\  are connected by
relations (2.7) satisfying \ $\buildrel\sim\over{f}\, (\lambda  \,, \,{\bf X} \,,  \,0)\
=\ \buildrel\circ\over{f}\, (\lambda  \,,  \, {\bf X})$ or equivalently \ $\buildrel\wedge\over{f}\, (t
 \,,  \,{\bf x}  \,,  \,0)\  =\ f\, (t  \,,  \, {\bf x})$). We then obtain
$$
\buildrel\sim\over{f}\, (\lambda  \,,  \,{\bf X}  \,,  \,\varepsilon) \
=\ \buildrel\wedge\over{f}\,\bigg( G\, (\lambda \,, \,{\bf X} \,, \,\varepsilon) \,, \,{\bf
\Phi}\,(\lambda \,, \,{\bf X} \,, \,\varepsilon) \,, \,\varepsilon \bigg)
\eqno{(2.11)}
$$
Let us
define Eulerian and Lagrangian variations of \ $f$
$$
\buildrel\wedge\over{\delta}f\  =\ {\partial \buildrel\wedge\over{f}   \over \partial
\varepsilon}\, (t \,, \, {\bf x} \,, \,0) \quad \hbox{ and } \quad  \buildrel\sim\over{\delta} f \
=\ {\partial \buildrel\sim\over{f} \over
\partial \varepsilon}\, (\lambda\,, \,{\bf X} \,, \,0)
$$
Differentiating relation (2.11) with respect to \ $\varepsilon$\
at \ $\varepsilon\  =\ 0$,   we get
$$
\displaystyle \buildrel\wedge\over{\delta}f\  =\ \buildrel\sim\over{\delta}f  \,- \,{\partial f\over
\partial {\bf z}}\, { \mbox{{\boldmath $\zeta$}}}
\eqno{(2.12)}
$$

\section{Governing equations}

Consider a four-dimensional vector \ ${\bf j}_0$\  satisfying
 conservation law
$$
{\rm Div} \, {\bf j}_0\ =\ 0
\eqno{(3.1)}
$$
Actually, (3.1) represents the mass conservation law, where \  ${\bf j}_0\,  =\, \rho
{\bf v},\, \rho$\  is the density and \ ${\bf v}^{*} \,  =\, (1, \,{\bf u}^{*})$\
is the four-dimensional
velocity vector. Let \ $a_k$\ be scalar quantities such as
the specific entropy, the number of bubbles per unit mass, the mass concentration, etc.,
which are conserved along the trajectories. Consequently, if
\ ${\bf j}_k\, =\,a_k {\bf j}_0$,
$$
{\rm Div} \,{\bf j}_k \ \equiv \ {\partial a_k\over \partial {\bf z}}\, {\bf j}_0 \  =\ 0,
\quad k \, =\, 1,{\dots},m
\eqno{(3.2)}
$$
Hence \ ${\bf j}_k$,\, $k\, =\, 1,\dots,m$\  form a set of solenoidal vectors collinear to \ ${\bf j}_0$.
Hamilton's principle {\it }needs the knowledge of Lagrangian of the medium. We take the
Lagrangian in the form
$$
L \ = \ \Lambda \bigg( {\bf j}_k  \,,  \,{\partial {\bf j}_k\over
\partial {\bf z}}
 \,,  \,\dots  \,,  \,{\partial^n {\bf j}_k\over \partial {\bf z}^n}  \,,  \,{\bf z} \bigg)
\eqno{(3.3)}
$$
 where \ ${\bf j}_k$\  are submitted to the constraints (3.1), (3.2) rewritten as
$$
{\rm Div}\,{\bf j}_k\  =\ 0,  \quad k\, =\, 0,\dots,m
\eqno{(3.4)}
$$
Let us consider three examples.

\noindent {\bf a -- Gas dynamics}Ê[1-3]

\noindent Lagrangian of the fluid is
\ $ \displaystyle L\,  =\, {1\over 2}\, \rho\, \vert {\bf u}\vert^2 \, -\, \varepsilon\, (\rho,
\,\eta) \,-\,\rho \,\Pi\,({\bf z})$, where \ $\varepsilon$\  is the
 internal energy per unit volume, \ $\eta \, =\, \rho s$\  is the entropy per unit volume, \ $s$\  is the
specific entropy and \ $\Pi$\  is an external potential. Hence, in variables
\ ${\bf j}_0 \, =\, \rho {\bf v}$,
\, ${\bf j}_1 \, =\,  \rho s {\bf v}$ , \, ${\bf z}$,  Lagrangian takes the form
$$
L\ =\ {1\over 2}\, \bigg( \,{ \vert {\bf j}_0\vert ^2\over
{\bf l}^{*} {\bf j}_0}\, -\, {\bf l}^{*} {\bf j}_0 \, \bigg) \, -\, \varepsilon\, ({\bf l}^{*} {\bf j}_0
\,,
\,{\bf l}^{*} {\bf j}_1) \,- \,{\bf l}^{*} {\bf j}_0 \, \Pi ( {\bf z}) \ = \
\Lambda \, ({\bf j}_0 \,, \,{\bf j}_1\,,\,{\bf z})
$$
 where \ ${\bf l}^{*} \, =\,(1,0,0,0)$.

\noindent {\bf b -- Thermocapillary fluids} [6,7]

\noindent Lagrangian of the fluid is
\ $ \displaystyle L\, =\,{1\over 2}\,\rho\,\vert {\bf u}\vert^2 \,-\,
\varepsilon\,\,(\rho \,,\,\nabla \rho \,, \,\eta \,, \,\nabla \eta) \,-\,\rho\, \Pi({\bf z})$.
Since \ $\nabla \rho\, =\,\nabla\, ( {\bf l}^{*} {\bf j}_0)$\  and \ $\nabla \eta\, =\,\nabla\,
( {\bf l}^{*} {\bf j}_1)$,  we obtain Lagrangian in the form (3.3)
$$
L\ =\ \Lambda\, \left( {\bf j}_0 \,, \,{\partial {\bf j}_0 \over \partial {\bf z}} \,,
\, {\bf j}_1  \,,  \,{\partial {\bf j}_1 \over \partial {\bf z}}  \,,  \,{\bf z} \right)
$$

\noindent {\bf c -- One-velocity bubbly liquids} [8-10]

$$
L\  =\ {1\over 2}\, \rho\, \vert {\bf u}\vert^2 \,-\, W \,
\bigg(\rho \,, \,{d \rho \over dt}\,, \,N \bigg), \quad \mbox{ with } \quad {d \rho \over dt} \
=\ {\partial \rho \over \partial {\bf z}}\, {\bf v}
$$
where \ $\rho$\  is now the average density of the bubbly liquid and \ $N$\  is the number of identical bubbles
per unit volume of the mixture. We define again
\ ${\bf j}_0\, =\,\rho {\bf v}$\  and \ ${\bf j}_1\, =\,N {\bf v}$.
By using \ $\displaystyle {d \rho \over dt}\, =\,{\partial \rho \over \partial {\bf z}}\, {\bf v}\,
=\, {\partial ( {\bf l}^{*} {\bf j}_0) \over \partial {\bf z}}\,{{\bf j}_0 \over {\bf l}^{*}  {\bf j}_0}$\
and \ $N\, =\,{\bf l}^{*} {\bf j}_1$,  we obtain Lagrangian in the form
$$
L\  =\ \Lambda\,\bigg( {\bf j}_0 \,, \,{\partial {\bf j}_0\over \partial {\bf z}}
\,, \,{\bf j}_1 \bigg)
$$

{\bf The Hamilton principle} reads:
{\it for each field of virtual displacements \ ${\bf z} \, \in \, \Omega \longrightarrow
{\mbox{{\boldmath $\zeta$}}}$\   such that \ ${ \mbox{{\boldmath $\zeta$}}}$\  and its derivatives are zero on}
\ $\partial \Omega$,
$$
\delta\, \int_\Omega\,\,\Lambda\,\, d \Omega \ =\ 0
\eqno{(3.5)}
$$
Since variation \ $\buildrel\wedge\over{\delta}$\  is independent of domain \ $\Omega$\
and measure \ $d\Omega$, the variation of Hamilton action (3.5) with
the zero boundary conditions for \ ${\bf j}_k$\  and its derivatives yields
$$
\delta \int_\Omega \Lambda \ d\Omega \ =\  \int_\Omega \sum^m_{k=0} \ {\delta \Lambda \over
\delta {\bf j}_k}\,\buildrel\wedge\over{\delta} {\bf j}_k\ d\Omega \ = \ 0
\eqno{(3.6)}
$$
where \ $\displaystyle\, {\delta \over \delta {\bf j}_k}$\  refers to the variational derivative with respect
to \ ${\bf j}_k$.  In particular, if Lagrangian (3.3) is
$$
L \ = \ \Lambda \, \bigg( {\bf j}_k \,, \,{\partial {\bf j}_k\over \partial {\bf z}} \,, \,
{\bf z} \bigg)
$$
we get
$$
{\delta\Lambda\over \delta{\bf j}_k}\  =\ {\partial \Lambda \over \partial {\bf j}_k}
\, -\, {\rm Div} \, \left(\,\,{\partial \Lambda\over \partial \pmatrix{\displaystyle{\partial {\bf j}_k\over
\displaystyle \partial {\bf z}}}} \right)
$$
We have to emphasize that in (3.6), the variations \ $\buildrel\wedge\over{\delta}{\bf j}_k$\
should take into
account constraints (3.4). We use the same method as in [2, p. 145] for the variation of density.
This method does not use Lagrange multipliers since the constraints (3.4) are
satisfied automatically. The calculation of \ $\buildrel\wedge\over{\delta}{\bf
j}_k$,\, $k\, =\,0,\dots,m$\  is performed in two steps.
 First, in  Appendix A we calculate Lagrangian variations
\ $\buildrel\sim\over{\delta}{\bf v}$, \, $\buildrel\sim\over{\delta}\rho$\  and \
$\buildrel\sim\over{\delta} a_k$\  (expressions (A.2), (A.5) and (A.6), respectively). Second, by using (2.12)
we obtain in Appendix B Eulerian variations
\ $\buildrel\wedge\over{\delta} {\bf j}_k \,, \,k\, =\, 0,\dots,m$\  (see (B.3))
$$
\buildrel\wedge\over{\delta} {\bf j}_k\  =\ \bigg( \, {\partial { \mbox{{\boldmath
$\zeta$}}}\over \partial {\bf z}} \,  -\, ({\rm Div} \,{ \mbox{{\boldmath $\zeta$}}} ) \,I \bigg)
\ {\bf j}_k \, -\, {\partial {\bf j}_k\over \partial {\bf z}}\,{ \mbox{{\boldmath $\zeta$}}}
$$
 Let us now define the four-dimensional covector
$$
{\bf K}_k^{*} \ =\ {\delta\Lambda\over \delta {\bf j}_k}
\eqno{(3.7)}
$$
Taking into account conditions (3.4) and the fact that \ ${ \mbox{{\boldmath $\zeta$}}}$\  and
its derivatives are zero on the boundary \ $\partial \Omega$,
equations (3.6), (B.3) yield
$$
\delta\int_\Omega\Lambda\,\,d\Omega\  = \ \int_\Omega\,\, \sum^m_{k=0}\,\, {\bf K}_k^{*}
\, \left( \left( {\partial  { \mbox{{\boldmath $\zeta$}}}\over
\partial {\bf z}} \ -\ ( {\rm Div} \,{ \mbox{{\boldmath $\zeta$}}})\, I \right) \,{\bf j}_k
\ -\ {\partial {\bf j}_k\over \partial {\bf z}} \,{ \mbox{{\boldmath $\zeta$}}} \right) \ d\Omega
$$
$$
= \  \int_\Omega\,\, \sum^m_{k=0}\,\, \left(  \,- \, {\rm Div} \,( {\bf j}_k \,{\bf K}^{*}_k)
\ +\ {\bf j}_k^{*}  \,{\partial {\bf K}_k\over \partial {\bf z}} \,\right)
\, { \mbox{{\boldmath $\zeta$}}} \ d\Omega
$$
$$
= \  \int_\Omega\,\, \sum^m_{k=0}\,\, {\bf j}_k^{*}\,\, \Bigg({\partial {\bf K}_k\over \partial {\bf z}}
\ -\ \bigg({\partial {\bf K}_k\over \partial {\bf z}}\bigg)^{*} \Bigg)\, { \mbox{{\boldmath $\zeta$}}}
\ d\Omega \ = \ 0
$$
Hamilton's principle yields the governing equations in the form
$$
\sum^m_{k=0}\,\, {\bf j}_k^{*} \, \Bigg( \,{\partial {\bf K}_k\over \partial {\bf z}}
\ -\ \bigg(\, {\partial {\bf K}_k\over \partial {\bf z}} \bigg)^{*} \Bigg) \  =\ 0
\eqno{(3.8)}
$$
 where \ ${\bf K}_k$\  are given by definition (3.7).
The system (3.4), (3.8) represents \ $m + d+1$\  partial differential equations for \ $m  + d $\
unknown functions \ ${\bf u}$,\, $\rho$,\, $a_k$,\, $k\, = \, 1,\dots,m$,
 where \ $d$\  is the dimension of the \ $\Omega$-space. Since the matrix
$$
R_k\  =\ {\partial {\bf K}_k \over \partial {\bf z}} \ -\ \bigg( \,{\partial {\bf
K}_k\over
\partial {\bf z}} \bigg)^{*}
\eqno{(3.9)}
$$
is antisymmetric and all the vectors \ ${\bf j}_k, \,k \, =\, 0,\dots,m$\  are
collinear, we obtain that \  ${\bf j}_k^{*}  \,R_k \,{\bf j}_0 \,\equiv \,0$.

\noindent Consequently
$$
\sum^m_{k=0} \, {\bf j}_k^{*} \, R_k \,{\bf j}_0 \ \equiv\ 0
\eqno{(3.10)}
$$
and the overdetermined system (3.4), (3.8) is compatible.

In the case \ $a_k\,  =\, {\rm const}, \,k\, =\,1,\dots,m$\  and \ $\displaystyle{
\Lambda\,  =\, \Lambda \left( {\bf j}_0, {\partial{\bf j}_0\over\partial {\bf z}}\right)}$, the system (3.4),
(3.8) can be rewritten in a simplified form
$$
\displaystyle \cases{ {\bf j}^{*}_0\ \Bigg( \displaystyle \ {\partial {\bf K}_0\over \partial {\bf z}}
\ -\ \bigg( {\partial {\bf K}_0\over \partial {\bf z}}\bigg)^{*} \Bigg) \  =\ 0\cr\cr
\displaystyle {\rm Div} \, {\bf j}_0 \  = \ 0}
\eqno(3.11)
$$
We call {\it potential motion} such solutions of (3.11) that
$$
{\bf K}_0^{*} \ = \ {\partial \varphi_0\over \partial {\bf z}}
\eqno{(3.12)}
$$
where \ $\varphi_0$\  is a scalar function.
In this case, the governing system for the potential motion is in the
form
$$
\cases{\displaystyle\ {\delta\Lambda\over \delta{\bf j}_0}\  =\ {\partial \varphi_0\over \partial {\bf
z}}\cr\cr
\hbox{Div }{\bf j}_0\  =\ 0}
\eqno{(3.13)}
$$
For the gas dynamics model, the system (3.12), (3.13) reads
$$
\cases{\displaystyle\ {\partial \varphi_0 \over \partial t}\  = \ - \pmatrix{\displaystyle {1\over
2}\,\vert {\bf  u}\vert ^2 \ +\ {\partial
\over \partial \rho}\, e(\rho) \,\,+\,\,\Pi}, \quad e(\rho) \  = \ \varepsilon\, (\rho , \rho s_{e})
\cr\cr
 \nabla\varphi _0\  =\ {\bf u}^{*} \cr\cr
\displaystyle\ {\partial \rho\over \partial t}\ +\  {\rm div}(\rho {\bf u})\  =\ 0}
$$
Eliminating the derivative $\displaystyle\, {\partial \varphi _0\over \partial t}$,
 we obtain the classical model for potential flows [2]
$$
\cases{\displaystyle\, {\partial {\bf u}^{*} \over \partial
t}\ +\ \nabla\, \pmatrix{\displaystyle\, {1\over 2}\vert {\bf u}\vert ^2 \ +\ {\partial \over
\partial \rho}\, e(\rho) \ +\ \Pi}\  =\ 0\cr\cr
 {\bf u}^{*} \  =\ \nabla \varphi_0\cr\cr
\displaystyle\, {\partial \rho\over \partial t}\ +\  {\rm div}(\rho {\bf u})\  =\ 0}
\eqno{(3.14)}
$$

\section{Conservation laws}

Equations (3.4), (3.8) can be rewritten in a divergence form. The demonstration
is performed for Lagrangian \ $\Lambda$\  which depends only on \ $\displaystyle\, {\bf j}_k \,,
\, {\partial {\bf j}_k\over
\partial {\bf z}} $\  and \ ${\bf z}$.
The following result is proved in Appendix C.

\bigskip

\noindent \underbar{Theorem 4.1}
\medskip

\noindent {\it Let \ $\displaystyle\, L\,  =\, \Lambda\, \bigg( {\bf j}_k  \,,  \,{\partial {\bf
j}_k\over \partial {\bf z}} \,, \,{\bf z}\bigg)$.  The following vector relation is an identity
$$
{\rm Div} \ \left( \sum^m_{k=0} \left( {\bf K}_k^{*} \,{\bf j}_k\,I \,-\, {\bf j}_k{\bf
K}_k^{*} \, +\, A^{*}_k \,{\partial {\bf j}_k\over \partial {\bf z}} \right) \, - \, \Lambda \,\,I
\right)
$$
$$
 + \,{\partial\Lambda\over\partial{\bf z}}
 \, + \,\sum^m_{k=0} {\bf K}_k^{*} \, {\rm Div} \,{\bf j}_k \,-\, \sum^m_{k=0} {\bf j}_k^{*}
 \,R_k \ \equiv\  0
\eqno(4.1)
$$
where the matrices \ $R_k$\  are given by (3.9) and}
\ $A^{*}_k\, =\, \displaystyle { \partial \Lambda\over \partial
\left(\displaystyle {{\partial {\bf j}_k\over\partial {\bf z}} } \right)}$.

\noindent In particular, we get

\bigskip

\noindent \underbar{Theorem 4.2}\medskip

\noindent {\it The governing equations (3.4), (3.8) are equivalent to the system of conservation laws}
$$
{\rm Div} \,\left(\sum^m_{k=0}\,\,\left( {\bf K}^{*} _k {\bf j}_k\,I \,-\,{\bf j}_k {\bf
K}_k^{*}
\,+\,A^{*}_k\,{\partial {\bf j}_k\over \partial {\bf z}}\right) \, - \,\Lambda I  \,\right)  \,+ \,{\partial
\Lambda\over \partial {\bf z}}\  =\ 0
$$
$$
{\rm Div} \,{\bf j}_k \  =\ 0, \quad k \, =\, 0,\dots,m
$$
In the special case of potential flows (3.13) the
governing equations admit additional conservation laws
$$
{\rm Div} \,\bigg( {\bf c }{\bf K}_0^{*} \,-\, ({\bf K}_0^{*}{\bf c }) I\bigg)\  =\ 0
\eqno(4.2)
$$
where \ ${\bf c}$\  is any constant vector. Indeed,
$$
{\rm Div} \, ({\bf c }{\bf K}_0^{*} )\  =\ {\bf c }^{*} \, \bigg(
{\partial {\bf K}_{0}
\over \partial {\bf z}}\bigg)^{*}, \quad
{\rm Div} \,({\bf K}_0^{*}{\bf c }\,\,I)\  =\ {\partial \over \partial {\bf z}}\, ({\bf K}_0^{*}{\bf c
}) \  =\ {\bf c }^{*} \, \bigg( {\partial {\bf K}_{0}\over \partial {\bf z}}\bigg)
$$
Since for the potential flows
$$
{\partial {\bf K}_0\over \partial {\bf z}}\  =\ \bigg( {\partial {\bf K}_0\over \partial
{\bf z } }\bigg)^{*}
$$
we obtain (4.2). In particular, if we take \ ${\bf c} = {\bf l}$ \ for the gas dynamics equations, we get
conservation laws (3.14).
Multiplying (4.1) by \ ${\bf j}_0$\  and taking into account the identity (3.10),
we obtain the following theorem

\bigskip

\noindent \underbar{Theorem 4.3} \medskip

\noindent
{\it The following scalar relation is an algebraic identity}
$$
\left( {\rm Div} \, \left(\sum^m_{k=0} \big( {\bf K}_k^{*} \,{\bf j}_k \,I \, -\, {\bf
j}_k{\bf K}_k^{*} \, +\, A^{*}_k\, {\partial {\bf j}_k\over \partial {\bf z}} \,\big ) \,-\,\Lambda I
\right) \right) \, {\bf j}_0
$$
$$
+\, {\partial \Lambda\over \partial {\bf z}}\,{\bf j}_0 \,+\,\sum^m_{k=0}\,\,({\bf
K}_k^{*} \, {\bf j}_0)\, {\rm Div} \,{\bf j}_k\ \equiv\  0
\eqno(4.3)
$$
This theorem is a general representation of the Gibbs identity expressing that the ''energy equation''
is a
consequence of the conservation of ''mass'', ''momentum'' and ''entropy''.
Examples of this identity for thermocapillary fluids and bubbly liquids were obtained previously in [6,10].

Identity (4.3) yields an important
consequence. Let us recall the Godunov-Friedrichs-Lax method of symmetrisation of quasilinear conservation laws [4-5]
(see also different applications and generalizations in [11-12]).
We suppose that the system of conservation laws for \ $n$\  variables \ ${\bf q}$\  has the form
$$
{\partial f_i\over \partial t}\, +\, {\rm div} \, {\bf F}_i \ =\ 0,   \quad  f_i\ =\ f_i\, ( {\bf q}),
\quad  {\bf F}_i \ =\ {\bf F}_i\, ( {\bf q}), \quad i=1,\dots,n
\eqno{(4.4)}
$$
Let us also assume that (4.4) admits an additional "energy" conservation law
$$
{\partial e\over \partial t}\ +\ {\rm div} \,  {\bf E}\  =\ 0, \quad e \ =\ e\, ( {\bf q}), \quad
{\bf E} \ =\ {\bf E}\, ( {\bf q})
$$
which is obtained by multiplying each equation of (4.4) by some functions \ $p^i$\  and
then by summing over \ $i\, = \, 1,\dots,n$
$$
{\partial e\over \partial t}\, + \, {\rm div}\,  {\bf E}\ \equiv\ p^i\left( {\partial
f_i\over \partial t}\, +\, {\rm div}\, {\bf F}_i \right)
\eqno{(4.5)}
$$
 In particular, if we consider \ $e$,\,
${\bf E}$\  and \ ${\bf F}_i$\  as functions of \ $f_i$,  we obtain from (4.5)
$$
{\partial e\over \partial f_i}\  =\ p^i, \quad {\partial {\bf E}\over \partial f_j} \
=\ p^i \,{\partial {\bf F}_i\over \partial f_j}\  =\ {\partial e\over
\partial f_i} \,{\partial{\bf F}_i\over \partial f_j}
\eqno{(4.6)}
$$
Let us introduce functions \ $N$\  and \ ${\bf M}$\  such that
$$
N\ =\ f_i \, p^i-e, \quad {\bf M}\ =\ {\bf F}_i p^i \, - \, {\bf E}
\eqno{(4.7)}
$$
Consequently, from equations (4.6) and (4.7) we find that
$$
{\partial N\over \partial p^i}\  =\ f_i, \quad {\partial {\bf M}\over \partial p^i}\  =\ {\bf
F}_i
\eqno{(4.8)}
$$
Hence, substituting (4.8) into (4.4), we get a {\it symmetric} system in the form
$$
{\partial^2N\over \partial p^i\partial p^j}\, {\partial p^j\over \partial t}
\, +\, tr\, \left( \,{\partial^2 {\bf M}\over \partial p^i\partial p^j} \,{\partial
p^j\over \partial {\bf x}}\right) \  =\ 0
\eqno{(4.9)}
$$
If matrix \ $\displaystyle\, N_{ij}\,
=\, {\partial^2N\over \partial p^i\partial p^j}$\  is positive definite then
 the symmetric system (4.9) is
 $t$-{\it hyperbolic symmetric in the sense of Friedrichs}.
Obviously, matrix $\displaystyle\, {\partial^2N \over \partial
p^i\partial p^j}$\  is positive definite if and only if matrix
\ $\displaystyle\, e^{ij}\,  =\, {\partial^2
e\over \partial f^i\partial f^j}$\  is positive definite, since
$e^{ij}N_{jp}\,  =\, \delta^i_p$,  where \ $\delta^i_p$\  is the
Kronecker symbol.

Now, let us rewrite identity (4.3) in the non-dispersive case
$$
\left( {\rm Div} \left( \sum^{m}_{k=0}
\left( {\bf K}^{*}_k {\bf j}_k \, I \, - \, {\bf j}_k {\bf K}^{*}_k \right)
\,-\, \Lambda\, I \right) \right) \, {\bf j}_0 \, +
\, \sum^{m}_{k=0} \left( {\bf K}^{*}_k \, {\bf j}_0 \right) {\rm div}\, {\bf j}_k\  \equiv\ 0
\eqno{(4.10)}
$$
Identity (4.10) is exactly of the same type as identity (4.5). It means that the system
(3.4), (3.8) can be always rewritten in a Godunov-Friedrichs-Lax symmetric form. Actually,
if we take
$$
e \ = \ \sum^{m}_{k=0} \, {\bf K}^{*}_k \, {\bf j}_k \, - \, ( {\bf j}_k \, {\bf K}^{*}_k )^1_1
\, -\, \Lambda \quad \mbox{ and } \quad {\bf E} \ = \ - P \, {\bf j}_k \, {\bf K} ^{*}
$$
where
$$
P \ = \
\pmatrix{
0 & 1 &  0 & 0 \cr
0 & 0 &  1 & 0 \cr
0 & 0 &  0 & 1 \cr
}
$$
the conjugate variables \ ${\bf p}$ \ are
$$
{\bf p} \ = \ \pmatrix{
{\bf u} \cr\cr
\displaystyle
- {1 \over \rho}\, {\bf K}^{*}_k \,{\bf j}_0 \cr\cr
}, \quad k \, = \, 0,\dots,m
$$
Consequently, the Gibbs identity (4.10) gives directly a set of conjugate variables. Therefore, we have proved the
following theorem

\bigskip

\noindent \underbar{Theorem 4.4} \medskip

\noindent {\it If \ $L \, =\,\Lambda ({\bf j}_k)$, the system (3.4), (3.8) is always
symmetrisable}.

\medskip

\noindent The property of convexity, needed for the hyperbolicity of the governing system, should be
verified for each particular case. For example, in the gas dynamics, the energy of the system is
given by the formula
$$
e\  =\ \sum^1_{k=0} \, {\bf K}^{*}_k\, {\bf j}_k \, -\, \sum^1_{k=0} \Big( {\bf j}_k \, {\bf K}^{*}
_k \Big)^1_1  \,-\, \Lambda \  =\ \varepsilon (\rho,
\eta)\, +\, \rho {\vert {\bf u} \vert^2 \over 2}
$$
Since
$$
{\bf K}^{*} _0 {\bf j}_0\  =\ \rho \left( {\vert {\bf u}\vert^2 \over 2}
\, -\, {\partial \varepsilon\over \partial \rho}\right), \quad
{\bf K}^{*}_1 \, {\bf j}_0\  =\ - \rho\, {\partial
\varepsilon\over \partial \eta}
$$
the conjugate variables are (see also [4])
$$
{\bf p}\  =\ \pmatrix{ {\bf u}\cr\cr
\displaystyle\,\,{\partial \varepsilon\over \partial \rho}\,\,-\,\,{\vert {\bf u}\vert ^2\over 2}\cr\cr
\displaystyle\,\,{\partial \varepsilon\over \partial \eta}} \quad = \quad
\pmatrix{ {\bf u}\cr\cr
\displaystyle\, \mu\, -\, {\vert {\bf u}\vert ^2\over 2}\cr\cr T}
$$
where \ $\mu$\  is the Gibbs potential and \ $T$\  is the temperature.
Obviously, if \ $\varepsilon (\rho,\eta)$\  is
convex, the total energy \ $e$\  is convex with respect to \ $\rho {\bf u}$,\, $\rho $\  and \ $\eta$.

\section{Stability of equilibrium states for potential flows.}

We assume that \ $L$\  is a function of \ ${\bf j}_k$\  and \ $\displaystyle\,{\partial {\bf j}_k\over
\partial {\bf z}}$\     besides it does not depend on \ ${\bf z}$, i.e.\
$$
L\  =\ \Lambda\, \left( {\bf j}_k  \,,  \,  {\partial {\bf j}_k\over
\partial {\bf z}}\right)
$$
Let us give some definitions. An {\it equilibrium state} is a solution of equations
(3.4), (3.8) such that \ ${\bf j}_k\,  = \,{\bf j}_{ke} \, =\, {\rm const}$.
Let \ ${ \mbox{{\boldmath $\nu$}}} $\  be a real unit vector.
Any vector \ ${ \mbox{{\boldmath $\beta$}}}$\  can be represented in the form
$$
{ \mbox{{\boldmath $\beta$}}}\  =\ \omega { \mbox{{\boldmath $\nu$}}} \, +\, { \mbox{{\boldmath
$\beta$}}}_{\sigma}, \quad \mbox{ where } \quad  {\mbox{{\boldmath $\nu$}}}^{*} \, { \mbox{{\boldmath
$\beta$}}}_{\sigma} \ =\ 0
$$
The equilibrium state \ $\displaystyle\, {\bf j}_{ke}\, \ne\, 0$\  is {\it
linearly stable in the direction} \ ${ \mbox{{\boldmath $\nu$}}} $\  if and only if all non-trivial solutions
of the form \ $ {\bf J}_k\, e^{i { \mbox{{\boldmath $\beta$}}}^{*}{\bf z}} $ of the system (3.4), (3.8)
linearized at the equilibrium state \ ${\bf j}_{ke}$\
are such that \ $\omega$\  is real for any real \ ${ \mbox{{\boldmath $\beta$}}}\sigma$.

\subsection{Non-dispersive case}

We note that in a non-dispersive case the stability means
hyperbolicity of governing system [13-14]. We omit the index ``0`` and rewrite system
(3.13) in the form
$$
\cases{ {\bf K}^{*} \ \equiv\ \displaystyle\, {\partial
\Lambda\over \partial {\bf j}}\  =\ \displaystyle\,{\partial \varphi \over \partial
{\bf z}}\cr\cr
{\rm Div}\,  {\bf j}\ =\ 0}
\eqno{(5.1)}
$$
where \ $\Lambda \, = \,\Lambda ({\bf j})$.
The Legendre transformation of \ $\Lambda ({\bf j})$ \ is
$$
\Delta ({\bf K})\  =\ {\bf K}^{*}
{\bf j}\ -\ \Lambda( {\bf j})
\eqno{(5.2)}
$$
If the matrix \ $\displaystyle\, \Lambda''( {\bf
j})\,  =\, \displaystyle\, {\partial \over \partial {\bf
j}}\, \left(\left(\,{\partial \Lambda\over \partial  {\bf j}}\right)^{*}
\right)$ \
 is non-degenerate, (5.2) involves the formula
$$
{\bf j}^{*} \  =\ {\partial \Delta\over \partial {\bf K}}
\eqno{(5.3)}
$$
Hence, relations (5.1) - (5.3) yield
$$
{\partial {\bf j}\over\partial {\bf z} }\  =\ {\partial \over \partial {\bf
K}}\, \left(\left(\,{\partial
\Delta\over \partial {\bf K}}\right)^{*} \right) \,{\partial {\bf K}\over\partial {\bf z} }\
=\ \Delta''({\bf K})\, {\partial {\bf K}\over\partial {\bf z} }\ =\ \Delta''({\bf K})\, \varphi
''({\bf z})\
\eqno{(5.4)}
$$
where
$$
\Delta''({\bf K})\  =\ {\partial \over \partial {\bf
K}}\, \left(\left( \,{\partial
\Delta\over \partial {\bf K}}\right)^{*} \right), \quad
\varphi'' ({\bf z}) \ = \ {\partial \over \partial {\bf
z}}\left(\left(\,{\partial
\varphi \over \partial {\bf z}}\right)^{*} \right)
$$
The second equation (5.1) and (5.4) involve
$$
tr \, \Big( \Delta''({\bf K})\, \varphi'' ({\bf z}) \Big) \ =\ 0
$$
If we replace
\ ${\bf \varphi}$ \ by \ $i \, \Phi e^{i { \mbox{{\boldmath $\beta$}}}^{*} {\bf z}}$, we get immediately the
{\it dispersion relation}
$$
{ \mbox{{\boldmath $\beta$}}}^{*} \Delta''({\bf K}_e){ \mbox{{\boldmath $\beta$}}}\
=\ 0
\eqno{(5.5)}
$$
where index "e" corresponds to the equilibrium state. The matrix
\ $\Delta''({\bf K}_e)$\  can be easily calculated in terms of Lagrangian
\ $\Lambda( {\bf j})$.
Indeed,
$$
I\ =\ {\partial {\bf j}\over \partial {\bf j}}\  =\ {\partial {\bf j}\over \partial {\bf
K}}\, {\partial {\bf K}\over \partial {\bf j}}\  =\ \Delta''({\bf K})\, \Lambda''({\bf j})
$$
It follows that
$$
\Delta''({\bf K}_e)\  =\  \Big(\Lambda''( {\bf j}_e)\Big)^{-1}
\eqno{(5.6)}
$$
Relations (5.5) - (5.6) imply the following result.

\bigskip

\noindent \underbar{Theorem 5.1} ({\it criterion of stability for potential non-dispersive
motions})\medskip

\noindent {\it If the symmetric matrix \ $\displaystyle\,G\, =\,\Lambda''( {\bf j}_e)$\
has the signature \ $(-,+,+,+)$, the equilibrium state \ ${\bf j}_e$\
is stable in any direction \ ${ \mbox{{\boldmath $\nu$}}}$\  belonging to the intersection
of two cones }
$$
C_1\ =\
\Bigl\{  {\mbox{{\boldmath $\nu$}}}
\left\vert \right.
{\mbox{{\boldmath $\nu$}}}^{*}
G^{-1}{\mbox{{\boldmath $\nu$}}} < 0 \Bigr\}
\quad \mbox{ and } C_2\ =\
\Bigl\{  {\mbox{{\boldmath $\nu$}}}
\left\vert \right.
{\mbox{{\boldmath $\nu$}}}^{*}
G{\mbox{{\boldmath $\nu$}}}<0  \Bigr\}
$$
\underbar{Proof}. Since \ $C_1$\  and \ $C_2$\  contain the
eigenvector corresponding to the negative eigenvalue of
\ $G$, $C_1\,\cap\,C_2\, \ne \,\emptyset$. We can find orthogonal coordinates
\ $(t, x_1,x_2,x_3)$\  such that the dispersion relation (5.5)
takes the form
$$
-t^2 \ +\ \sum^3_{i=1} \, \lambda_ix^2_i\ =\ 0, \quad \mbox{ with }
\quad \lambda_i>0
$$
In coordinates \ $(t,x_1, x_2,x_3)$ \ the cone \ $C_2$\  is defined by the inequality
$$
-t^2 \, +\, \sum^3_{i=1} \, {x^2_i\over \lambda_i} \ < \ 0
$$
Let us represent \ ${ \mbox{{\boldmath $\beta$}}}$ \ in the form
$$
{ \mbox{{\boldmath $\beta$}}} \ =\  \pmatrix{t\cr x_1\cr x_2\cr x_3}\, =\,\omega
\pmatrix{1\cr n_1\cr n_2\cr
n_3} \,+\,\pmatrix{-n_1y_1-n_2y_2-n_3y_3\cr y_1\cr y_2\cr y_3}
$$
The dispersion relation
\ ${ \mbox{{\boldmath $\beta$}}}^{*} G^{-1} { \mbox{{\boldmath $\beta$}}}\, =\,0$\
implies
$$
-\left( \sum^3_{i=1} \, n_iy_i-\omega\right)^2 \, +\, \sum^3_{i=1}\, \lambda_i\,(\omega n_i+y_i)^2\
=\  0
$$
It involves
$$
\omega^2 \left( -1 \, +\, \sum^3_{i=1} \, \lambda_in^2_i\right) \, +\, 2\omega
\, \sum^3_{i=1}\, (1\, +\, \lambda_i)\, n_i y_i \,+ \,\sum^3_{i=1}\, \lambda_iy^2_i
\, -\, \left( \sum^3_{i=1}\, n_i y_i\right)^2 \ =\ 0
$$
Due to the following inequality,
$$
\sum^3_{i=1}\, \lambda_i y_i^2  \,- \,\left( \sum^3_{i=1} \, n_i y_i\right)^2 \ = \
\sum^3_{i=1} \, \lambda_i y_i^2
\, - \, \left( \sum^3_{i=1}\,{n_i\over
\sqrt{\lambda_i}} \,\sqrt{\lambda_i}\,\,y_i\right)^2
$$
$$
\ge\ \sum^3_{i=1}\, \lambda_i y_i^2 \, -\, \left( \sum^3_{i=1}\, {n^2_i\over
\lambda_i}\right)\, \left( \sum^3_{i=1}\, \lambda_iy_i^2\right) \
=\ \left( \sum^3_{i=1} \, \lambda_i y_i^2\right) \,\left( 1 \, -
\sum^3_{i=1} \,{n^2_i\over \lambda_i}\right)
$$
$\omega $ \ is real if \ ${ \mbox{{\boldmath $\nu$}}}$ belongs simultaneously to \ $C_1$ \ and \ $C_2$. The
theorem is proved.

For the gas dynamics model, if the volume energy is a convex
function of \ $\rho$\  and \ $\eta\,  =\, \rho s$, the matrix \ $G$\  satisfies
the conditions of Theorem 5.1.

\subsection{Dispersive case}
If \ $\displaystyle\, L\,
=\, \Lambda \pmatrix{ {\bf j}, \displaystyle\,\,{\partial {\bf j}\over
\partial {\bf z}}}$,
the governing system is
$$
\cases{ {\bf K}^{*} \ \equiv\ \displaystyle\, {\delta\Lambda\over \delta {\bf j}}\
=\ {\partial\Lambda\over
\partial {\bf j}} \, -\, {\rm Div} \pmatrix{\displaystyle\, {\partial \Lambda\over \partial
\pmatrix{\displaystyle\, {\partial {\bf j}\over \partial
 {\bf z}}}}}\  =\ \displaystyle\, {\partial
\varphi \over \partial {\bf z}}\, \cr\cr {\rm Div} \, {\bf j}\  =\ 0}
$$
The linearised system in a coordinate form is
$$
\Lambda_{ks}j^s \, +\, \Gamma^p_{ks}j^s_{,p} \, -\, { D}^{mp}_{ks}j^s_{,mp}\
=\ \varphi,_k,
\quad \mbox{ and } \quad j^s_{,s}\ =\  0
\eqno{(5.7)}
$$
where the comma denotes the derivative with respect to
\ $z^k$\  while \ $\Lambda_{ks}$,\, $\Gamma^p_{ks}$,\,
${ D}^{mp}_{ks}$\ , calculated at point \ ${\bf j}_e$, are defined by the relations
$$
\Lambda_{ks}\  =\  {\partial^2\Lambda\over \partial j^k\partial j^s}, \quad
\Gamma^p_{ks}\ =\ {\partial ^2\Lambda\over \partial j^k \partial j^s_{,p}}\,
-\, {\partial ^2\Lambda\over \partial j^s\partial j^k_{,p}}, \quad
{D}^{mp}_{ks}\ =\
{\partial ^2\Lambda\over \partial j^k_{,m}\,\partial j^s_{,p}}
\eqno{(5.8)}
$$
The following symmetry relations result from (5.8)
$$
\Lambda_{ks}\ =\ \Lambda_{sk}, \quad \Gamma^p_{ks}\ =\ -\Gamma^p_{sk},
\quad {D}^{mp}_{ks} \  =\ { D}^{pm}_{sk}
\eqno{(5.9)}
$$
For the solution of (5.7) in the form
\ $\displaystyle j^s\,  =\, J^s \, e^{i\beta_p z^p}$,\,
$\varphi \, =\,i \,\Phi \,
e^{i\beta_p z^p}$,  we get the following dispersion relation
$$
\det \pmatrix{\displaystyle G&& { \mbox{{\boldmath $\beta$}}}\cr\cr
{ \mbox{{\boldmath $\beta$}}}^{*} &&0}\ \equiv\ -{ \mbox{{\boldmath $\beta$}}}^{*}
\, {\rm Adj} \, (G) { \mbox{{\boldmath $\beta$}}}\  =\ 0
\eqno{(5.10)}
$$
where \ ${\rm Adj} \, (G)$\  is the adjoint matrix to \ $G$,  and the elements \ $G_{ks}$\
of the matrix \ $G$\  are defined by
$$
G_{ks}\ =\ \Lambda_{ks} \, +\, i\, \Gamma^p_{ks}\beta_p  \,+ \,{ D}^{mp}_{ks}\beta_m\beta_p
\eqno{(5.11)}
$$
Equations (5.9) - (5.11) imply that \ $G$\  is {hermitian} matrix.
In particular, if \ $\det G\, \ne \,0 $\  then \ ${\rm Adj} \,(G)\,  =\, \det (G)\, G^{-1}$\  and the
dispersion relation (5.10) is equivalent to
 \ ${ \mbox{{\boldmath $\beta$}}}^{*} G^{-1}{ \mbox{{\boldmath $\beta$}}}\,  =\, 0$,
which is a generalization of (5.5) for the dispersive case. We get the following obvious result:

\bigskip

\noindent \underbar{{Theorem 5.2}}\,\,({\it criterion of stability for potential dispersive motions})
\medskip

\noindent {\it If \ $ \Gamma^p_{ks} \, = \, 0$ and the symmetric matrix \ $G$ \ defined by (5.8), (5.11)
has the signature \ $(-,+,+,+)$ \ for \ ${\mbox{{\boldmath $\beta$}}} = 0$, the equilibrium state
 \ ${\bf j}_e$\  is stable for small \ ${ \mbox{{\boldmath $\beta$}}}$\  in any direction
\ ${ \mbox{{\boldmath $\nu$}}}$\  belonging to the intersection of the cones \ $C_i$,\, $i=1,2$ \ defined
in theorem 5.1.}

\medskip

\noindent We note that \ $\Gamma^p_{ks}$ \ defined by (5.8) are always zero
 if the expansion of Lagrangian \ $\Lambda$\
in Taylor series at the vicinity of equilibrium state does not contain linear terms with
respect to \ $\displaystyle {\partial {\bf j}_k \over \partial {\bf z}}$.

\noindent Sometimes, we are able to obtain a ``global'' stability. For example, let us
consider a particular case of bubbly liquids, for \ $N/\rho \, = \, {\rm const}$,
 defined in Section 3
$$
L\ =\ {1\over 2}\, \rho\vert {\bf u}\vert ^2 \, +\, {a\over 2} \left(\,{d\rho\over
dt}\right)^2 \, -\, \varepsilon(\rho)
$$
where \ $\varepsilon(\rho)$\  is a convex function of the density
and
\ $a$\  is a positive function of \ $\rho$. Then,
$$
L\  =\ {1\over 2} \left( \,{\vert {\bf j}\vert ^2\over {\bf l}^{*}
{\bf j}} \,-\,{\bf l}^{*}  {\bf j} \right) \,+\,
{a\over 2}\left( \,{\partial \over
\partial {\bf z}}({\bf l}^{*} {\bf j})\, { {\bf j}\over {\bf l}^{*} {\bf j}}\right)^2 \, -\, \varepsilon
({\bf l}^{*} {\bf j})
$$
Since the governing equations are invariant with respect to the Galilean
transformation
$$
t'\  =\ t, \quad
 {\bf x}'\  =\ {\bf x}\, +\, {\bf U} t, \quad
 {\bf u}'\  =\ {\bf u}\, +\, {\bf U}
$$
 we can always assume that ${\bf u}_e \, =\,0$\
(it is sufficient to consider the governing system in the reference frame moving with the velocity
\ ${\bf U}\, =\, {\bf u}_e $). Hence,
 \ $ {\bf j}_e\,  =\, \rho_e\, {\bf l}$, where ${\bf l}^{*} \, =\,(1,0,0,0)$.
Omitting index ''$e$'',  we can calculate the matrix \ $G$ defined by (5.11)
$$
 G\ =\ \pmatrix{\displaystyle -\, {\partial ^2\varepsilon\over \partial \rho^2}
\, +\, a\beta^2_1 & {\bf 0}^{*}
\cr
 {\bf 0}  &  \displaystyle\,{I \over \rho}\cr }
$$
Hence,
$$
{\rm Adj} \, (G)\  =\ \pmatrix{\displaystyle  {1\over \rho^3} & {\bf 0}^{*}
\cr
 {\bf 0} & \displaystyle\, {\left( a\beta^2_1
\, -\, {\partial ^2\varepsilon\over \partial
\rho^2}\right)  \,I \over \rho^2}\cr}
$$
and dispersion relation (5.10) reads
$$
{\beta^2_1\over \rho^3}\, +\, {a\beta^2_1 \, -\, \displaystyle\, {\partial ^2\varepsilon\over
\partial \rho^2}  \over \rho^2}\, \Big(\beta^2_2 \, +\, \beta^2_3 \, +\, \beta^2_4 \Big)\
=\ 0
\eqno{(5.12)}
$$
Let \ ${ \mbox{{\boldmath $\nu$}}}$\  be the direction of  time in
the time-space, then \ $\displaystyle\, { \mbox{{\boldmath
$\nu$}}}\,  =\, {\bf l}$\ and (5.12) is equivalent to
$$
{\omega^2\over   \displaystyle{\partial^2\varepsilon\over \partial \rho^2} \, -\, a \omega^2 }\
=\ \rho\ \vert {\mbox{{\boldmath $\beta$}}}_\sigma\vert ^2
\eqno{(5.13)}
$$
The graph of the dispersion
relation (5.13) for positive values of \ $\omega$\  is presented in Figure 1. We have
 denoted by \ $\displaystyle\, \omega_{*} \,  = \, \sqrt{ {\partial ^2\varepsilon\over \partial \rho^2}
\, {1\over a }}$\  the {\it eigenfrequency} of bubbles and by \ $c_0 \, = \, \sqrt{\displaystyle\, \rho
{\partial ^2\varepsilon\over \partial \rho^2}}$\  the {\it equilibrium sound speed} of bubbly liquid (see
[8]).
\begin{figure}[h]
\begin{center}
\includegraphics[width=9cm]{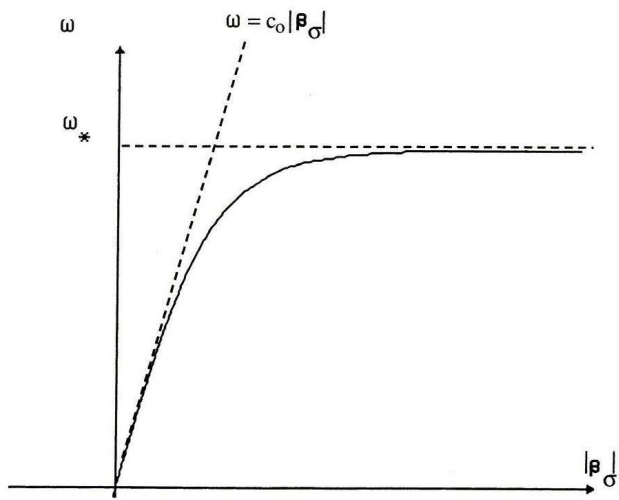}
\end{center}
\caption{Dispersion relation for bubbly liquids} \label{fig1}
\end{figure}

\section{Governing equations for mixtures}

Let us consider homogeneous binary mixtures. The mixture is described by the velocities \ ${\bf u}_{\alpha}$,
the average densities \ $\rho_{\alpha}$ \ and the specific entropies \ $s_{\alpha}$ \ for each component
($\alpha = 1, 2$). We introduce two reference frames associated with each component in the form
$$
\displaystyle \cases{t\  =\ g_\alpha(\lambda_\alpha  \,,  \,{\bf X}_\alpha)\cr\cr
{\bf x}\  =\ { \mbox{{\boldmath $\phi$}}}_\alpha(\lambda_\alpha   \,,  \,{\bf X}_\alpha)}
\eqno{(6.1)}
$$
and the inverse mappings
$$
\displaystyle \cases{\lambda_\alpha\  =\ h_\alpha(t  \,,  \,{\bf x})\cr\cr
{\bf X}_\alpha\  =\ {\bf \Psi}_\alpha(t  \,,  \,{\bf x})}
\eqno{(6.2)}
$$
The corresponding families of virtual motions generated by (6.1) are defined by
$$
\displaystyle \cases{t\  =\ G_\alpha(\lambda_\alpha  \,,  \,{\bf X}_\alpha  \,,
\,\varepsilon_\alpha)\cr\cr
{\bf x}\  =\ {\bf \Phi}_\alpha(\lambda_\alpha  \,,  \,{\bf X}_\alpha  \,,  \,\varepsilon_\alpha)}
$$
with
$$
\displaystyle \cases{G_\alpha(\lambda_\alpha  \,,  \,{\bf X}_\alpha  \,,  \,0)\
=\ g_\alpha(\lambda_\alpha  \,,  \,{\bf X}_\alpha)\cr\cr
 {\bf \Phi}_\alpha(\lambda_\alpha  \,,  \,{\bf X}_\alpha  \,,  \,0)\  =\ { \mbox{{\boldmath
$\phi$}}}_\alpha(\lambda_\alpha  \,,  \,X_\alpha)}
$$
We define the two Eulerian displacement
\ ${\mbox{{\boldmath $\zeta$}}}_\alpha\,  =\, ( {\tau}_\alpha  \,,  \,{ \mbox{{\boldmath
$\xi$}}}_\alpha)$\  where
$$
{\tau}_\alpha\  =\ {\partial G_\alpha\over \partial \varepsilon_\alpha}\, (\lambda_\alpha
 \,,  \,{\bf X}_\alpha  \,,  \,0), Ê\quad { \mbox{{\boldmath $\xi$}}}_\alpha\  =\ {\partial
{\bf \Phi}_\alpha\over \partial
\varepsilon_\alpha} \,(\lambda_\alpha  \,,  \,{\bf X}_\alpha  \,,  \,0)
$$
As in Section 3 we define tensor
 quantities associated with the two virtual
motions and variations \ $\buildrel\wedge\over\delta_\alpha$\  and
\ $\buildrel\sim\over\delta_\alpha$. In the
general case, we have two four-dimensional solenoidal vectors
\ ${\bf j}_{0(\alpha)}\,
=\, \rho_\alpha\, {\bf v}_\alpha$ \ corresponding to the \ $\alpha^{\rm th}$\  component,
where \ $\rho_\alpha$\  is the density and
\ ${\bf v}^{{*}}_{\alpha} \, =\,  (1,{\bf u}^{{*}}_{\alpha })$
 is the four-dimensional velocity vector.
 As in Section 3, we introduce additional physical quantities \ $a_{ k(\alpha)} $\
 associated with the four-dimensional vectors
\ ${\bf j}_{ k(\alpha)}\,  =\, a_{ k(\alpha)} \, {\bf j}_{0(\alpha)}$,\,  $\alpha\,  =\, 1,2$,\,
${k(\alpha)} \, =\,1,\dots,m(\alpha)$\  submitted to the constraints
$$
{\rm Div} \, {\bf j}_{ k(\alpha)}\  =\ 0
\eqno{(6.3)}
$$
For Hamilton's action in the form
$$
 a\ =\ \int_\Omega\,\,\Lambda\,\,\bigg( {\bf j}_{ k(\alpha)}  \,,  \,{\partial {\bf j}_{ k(\alpha)}\over
\partial {\bf z}}  \,,  \,\dots  \,, \, {\partial^n {\bf j}_{ k(\alpha)}\over
\partial {\bf z}^n} \,,  \,{\bf z}\bigg) \, d\Omega
$$
{\bf the Hamilton principle} reads:
{\it for each field of virtual displacements
\ ${\bf z} \, \in \, \Omega\longrightarrow { \mbox{{\boldmath $\zeta$}}}_{\alpha}$\
such that \ ${ \mbox{{\boldmath $\zeta$}}}_{\alpha}$\  and its derivatives are zero on}
\ $\partial \Omega $,
$$
\delta_\alpha \int_\Omega\,\, \Lambda\,\, d \Omega\  =\ 0
$$
Here \ $\delta_\alpha \,a$\  is the derivative of the Hamilton action with respect to \ $\varepsilon_\alpha$,
and \ ${ \mbox{{\boldmath $\zeta$}}}_{\alpha}$\  are the virtual  displacements expressed in
Eulerian coordinates
\ ${\bf z}$ \ by means of equations (6.2).
 The method developed in Section 3 yields the equations of motion in the form
$$
\sum^{ m(\alpha)}_{ k(\alpha) = 0}\,\, {\bf j}_{ k(\alpha)}^{*} \, \Bigg(\, {\partial {\bf
K}_{ k(\alpha)}
\over \partial {\bf z}} \, - \, \bigg({\partial {\bf
K}_{ k(\alpha)}
\over \partial {\bf z}}\bigg)^{*} \Bigg)\  =\ 0, \quad {\bf K}_{ k(\alpha)}^{*} \
=\ {\delta\Lambda\over
\delta {\bf j}_{ k(\alpha)}}
\eqno(6.4)
$$
$$
{\rm Div}\, {\bf j}_{ k(\alpha)}\  =\ 0, \quad \alpha\   = \ 1,2 \quad \mbox{ and } \quad k(\alpha) \ = \
0,\dots,m(\alpha)
$$
As in Theorem 4.1, for the case \ $\displaystyle \Lambda\,
=\, \Lambda\, \bigg( {\bf j}_{ k(\alpha)}  \,,  \,{\partial {\bf j}_{ k(\alpha)}
\over \partial {\bf z}}  \,,  \,{\bf z}\bigg)$ we can obtain the following identity
$$
{\rm Div}\, \left(  \sum^2_{\alpha=1} \, \sum^{ m(\alpha)}_{{ k(\alpha)} =0} \,\,\left( {\bf K}^{*}
_{ k(\alpha)} {\bf j}_{ k(\alpha)}\, I \, -\, {\bf j}_{ k(\alpha)}{\bf K}^{*}
_{ k(\alpha)} \, +\, A^{*}_{ k(\alpha)} \, {\partial {\bf j}_{ k(\alpha)} \over \partial {\bf
z}}\right)\, -\, \Lambda \, I\right)
$$
$$
+\,  {\partial \Lambda\over \partial {\bf z}} \, +\,  \sum^2_{\alpha=1}\, \sum^{
m(\alpha)}_{k(\alpha) =0}\, {\bf K}^{*} _{ k(\alpha)}\, {\rm Div}\, {\bf j}_{k(\alpha)}
\, -\, \sum^2_{\alpha=1} \, \sum^{ m(\alpha)}_{{ k(\alpha)} =0}\, {\bf j}_{ k(\alpha)}^{*}\, R_{
k(\alpha)}\ \equiv\ 0
$$
where
$$
A^{*}_{ k(\alpha)}\  =\ \displaystyle{\partial \Lambda\over \partial \left(\displaystyle{\partial
{\bf j}_{ k(\alpha)}\over
\partial {\bf z}}\right)} \quad \mbox{ and } \quad  R_{ k(\alpha)}\  =\ {\partial {\bf K}_{
k(\alpha)}\over \partial {\bf z}} \, -\, \bigg( {\partial {\bf K}_{ k(\alpha)}\over \partial {\bf z}}
\bigg)^{*}
$$
Hence, the governing equations (6.4)  admit the conservation laws
$$
{\rm Div}\, \left(  \sum^2_{\alpha=1}\, \sum^{m(\alpha)}_{{k(\alpha)} =0}\, \left( {\bf K}^{*}
_{k(\alpha)} {\bf j}_{k(\alpha)}I  \,-\, {\bf j}_{k(\alpha)}{\bf K}^{*}
_{k(\alpha)} \, +\, A^{*}_{k(\alpha)} \, {\partial {\bf j}_{k(\alpha)} \over \partial {\bf
z}}\right)\, -\, \Lambda\,I\right)  \,+\, {\partial \Lambda\over \partial {\bf z}}\  =\ 0
$$
$$
{\rm Div}\, {\bf j}_{k(\alpha)}\  =\ 0, \quad \alpha\  =\ 1,2 \quad \mbox{ and } k(\alpha)\
=\ 0,\dots,m(\alpha)
$$
We notice that for a one-velocity model, the number of scalar conservation laws is
\ $m \, +\, d \, +\, 1$\  where \ $d$\  is the dimension of the time-space. The number of unknown variables
is
\ $m \, +\, d$. Due to Theorem 4.3, the \ $m \, +\, d \, +\, 1$\  conservation laws are connected by the
''Gibbs identity''. In the case of mixtures, we obtain \ $\displaystyle\sum_{\alpha} m(\alpha) \, +\, d
\, +\, 1$\  conservation laws for \ $\displaystyle\sum_{\alpha} m(\alpha) \, +\, 2d$\  unknown variables.
In the general case, the classical approach based on
 conservation laws, does not allow to obtain Rankine-Hugoniot conditions for this system.
Nevertheless, in [15-17] we have obtained the jump conditions. The Hamilton principle
provides these conditions without any ambiguity.

\bigskip

\section{Linear stability of mixtures.}

 We consider the Lagrangian in
the form [15-19]
$$
L\, =\, {1\over 2}\, \rho_1\, \vert {\bf u}_1\vert^2 \, +\, {1\over
2}\, \rho_2\, \vert {\bf u}_2\vert^2 \, -\, W\,\,(\rho_1  \,,  \,\rho_2  \,,  \,\eta_1  \,,
\, \eta_2  \,,  \,{\bf w})
\eqno{(7.1)}
$$
where \ ${\bf w}\,  =\, {\bf u}_2 \, -\, {\bf u}_1$\  and \
$\eta_\alpha$\  is  the entropy per unit volume of the \ $\alpha^{th}$\  component. A generalisation of (7.1)
for thermocapillary fluids was also proposed
 in [20].
If \ $W$ \ is an isotropic function of \ ${\bf w}$, Lagrangian (7.1) can be rewritten as
follows
$$
\Lambda\  =\ {1\over 2}\, \sum^2_{\alpha\,  = \,1} \, \bigg(\, { \vert
{\bf j}_{0(\alpha)}\vert^2 \over {\bf l}^{*}\,\,{\bf j}_{0(\alpha)}} \,-\, {\bf
l}^{*}\, {\bf j}_{0(\alpha)}\bigg)
$$
$$
\, - \, W\, \bigg( {\bf l}^{*}\, {\bf j}_{0(1)}  \,,  \,{\bf
l}^{*} \,{\bf j}_{0(2)}  \,,  \,{\bf l}^{*} \,{\bf j}_{1(1)}  \,,  \,{\bf l}^{*} \,{\bf j}_{1(2)} \,,
\,\mu \bigg)
\eqno{(7.2)}
$$
where
$$
\mu \ =\ {1\over 2} \left\vert  {{\bf j}_{0(2)} \over {\bf l}^{*}{\bf j}_{0(2)}
}\, -\, {{\bf j}_{0(1)} \over {\bf l}^{*}{\bf j}_{0(1)} }\right\vert^2
\eqno{(7.3)}
$$
Here we shall
also restrict our study to the case \ $s_{\alpha}\,  = \, {\rm const}$. To avoid double indices, we will
denote
\ ${\bf j}_{0(1)}$,\,
${\bf j}_{0(2)}$\  by \ ${\bf j}_1$,\, ${\bf j}_2$.
Hence, (7.2) and (7.3) involve
$$
\Lambda ({\bf j}_1,{\bf j}_2)\  =\ {1\over 2}\, \sum^2_{\alpha=1}\, \left(\,{\vert {\bf
j}_\alpha
\vert^2\over {\bf l}^{*} {\bf j}_\alpha}\, -\, {\bf l}^{*} {\bf j}_\alpha\right)  \, -\, {\buildrel\sim\over
W}({\bf l}^{*} {\bf j}_1  \,,
  \,{\bf l}^{*} {\bf j}_2  \,,  \,\mu)
\eqno{(7.4)}
$$
where \ ${\buildrel\sim\over W}(\rho_1,\rho_2,\mu)\
=\ W(\rho_1,\rho_2,\rho_1s_1,\rho_2s_2,\mu)$.

\noindent
Omitting the tilde symbol, we consider the potential motions
$$
\cases{ \displaystyle {\bf K}_\alpha^{*} \ \equiv\ {\partial \Lambda\over \partial {\bf j}_\alpha}\
=\ {\partial \varphi _\alpha\over \partial  {\bf z}}\cr\cr
\displaystyle {\rm Div}\, {\bf j}_\alpha\  =\ 0, \quad  \alpha=1,2}
\eqno{(7.5)}
$$
Let us consider the Legendre transformation of \ $\Lambda ({\bf j}_1,{\bf j}_2)$\
$$
\Delta( {\bf K}_1  \,,  \,{\bf K}_2)\  =\ \sum^2_{\alpha=1} \, {\bf j}_\alpha^{*} {\bf K}_\alpha
\, -\, \Lambda
\eqno{(7.6)}
$$
If the matrices
$$
\Lambda_{(\alpha \beta)}\ \equiv \ {\partial \over \partial {\bf
j}_\alpha}\left( \left(\,{\partial \Lambda\over
\partial {\bf j}_\beta}\right)^{*} \right) \  = \ \Lambda^{*}_{(\beta \alpha)}
\eqno{(7.7)}
$$
are non-degenerate, relation (7.6) involves
$$
{\bf j}^{*} _\alpha\  =\ {\partial \Delta\over \partial {\bf K}_\alpha}
\eqno{(7.8)}
$$
Substituting (7.8) in \ ${\rm Div} \, {\bf j}_\alpha\, =\,0$,  we get
$$
tr\left(\,{\partial {\bf j}_\alpha\over \partial {\bf z}}\right)\
=\ tr\left(\,{\partial \over \partial {\bf z}}\left( \,{\partial
\Delta\over \partial {\bf K}_\alpha}\right)^{*}\right)
$$
$$
=\ tr\left( \sum^2_{\gamma=1}\, {\partial \over
\partial {\bf K_\gamma}}\left( \,{\partial \Delta\over \partial {\bf K}_\alpha}\right)^{*}
\,{\partial {\bf K}_\gamma\over
\partial {\bf z}}\right) \   =\ tr\left( \sum_{\gamma=1}^2 \, \Delta_{(\gamma\alpha)}\varphi
''_\gamma\right)
\  =\ 0
$$
with
$$
\Delta_{(\gamma\alpha)}\ \equiv \ {\partial \over \partial {\bf
K}_\gamma}\left( \left(\,{\partial \Delta\over
\partial {\bf K}_\alpha}\right)^{*} \right) \  = \ \Delta^{*}_{(\alpha \gamma)}
\eqno{(7.9)}
$$
Hence, the equations of potential motions are
$$
tr\left( \sum^2_{\gamma=1} \, \Delta_{(\gamma\alpha )} \varphi ''_\gamma\right) \  =\ 0
\eqno{(7.10)}
$$
In the equilibrium state \ $\Delta_{(\gamma\alpha)}\, = \, {\rm const}$. Substituting
\ $\varphi_\gamma$ \  by \ $i
\Phi_\gamma e^{i{ \mbox{{\boldmath $\beta$}}}^{*} {\bf z}}$\  in (7.10),
we obtain the dispersion relation in the symmetric form
$$
\det\,\pmatrix{ { \mbox{{\boldmath $\beta$}}}^{*} \Delta_{(11)}{ \mbox{{\boldmath $\beta$}}}Ê&
{ \mbox{{\boldmath $\beta$}}}^{*}
\Delta_{(21)}{ \mbox{{\boldmath $\beta$}}}\cr { \mbox{{\boldmath $\beta$}}}^{*}\Delta_{(12)}{
\mbox{{\boldmath $\beta$}}} &{ \mbox{{\boldmath $\beta$}}}^{*}
\Delta_{(22)}{ \mbox{{\boldmath $\beta$}}}\cr }
\ =\ 0
$$
\noindent which is the generalisation of the dispersion relation (5.5). However, to calculate
\ $\Delta_{(\gamma\alpha)}$\  in terms of
\ $\Lambda_{(\gamma\alpha)}$\  defined by (7.7), (7.9)), we have to solve
the following system of matrix equations
$$
\sum^2_{\gamma=1}\, \Delta_{(\gamma\alpha)}\Lambda_{(\beta\gamma)} \  =\ I \,\delta_{\alpha\beta}, \quad
\delta_{\alpha\beta}\  =\ \cases{1,\quad \alpha=\beta\cr 0, \quad \alpha\ne \beta}
$$
A simpler method is to consider the linearised system generated from (7.5)
$$
\sum^2_{\gamma=1}\, \Lambda_{(\gamma\alpha)} {\bf j}_\gamma \, -\, \left( {\partial \varphi
_\alpha\over\partial {\bf z}}\right)^{*} \ =\ 0, \quad {\rm Div}\, {\bf j}_\alpha\ =\ 0
$$
where
\ $\Lambda_{(\gamma\alpha)}$\  are taken at the equilibrium state
\ ${\bf j}_{ke}$.  Substituting
\ ${\bf j}_\gamma$\  by \ ${\bf J}_\gamma\, e^{i { \mbox{{\boldmath $\beta$}}}^{*} {\bf
z}}$ \ and
\ $\varphi_\alpha$\  by \ $i\Phi_\alpha  \,e^{i { \mbox{{\boldmath $\beta$}}}^{*} {\bf
z}}$, we obtain
 the dispersion relation in the symmetric form
$$
\det\ \
\pmatrix{\Lambda_{(11)} & \Lambda_{(21)} &
{\mbox{{\boldmath $\beta$}}} & {\bf 0}\cr
\Lambda_{(12)} & \Lambda_{(22)} & {\bf 0} & { \mbox{{\boldmath $\beta$}}}\cr
 { \mbox{{\boldmath $\beta$}}}^{*} & {\bf 0}^{*} & {0} & {0}\cr
{\bf 0}^{*} &  {\mbox{{\boldmath $\beta$}}}^{*} & {0} & { 0}\cr}
 \ = \ 0
\eqno{(7.11)}
$$
Equation (7.11) is presented in terms of the matrices \ $\Lambda_{(\gamma\alpha)}$\
calculated directly from the Lagrangian.
Let us consider Lagrangian (7.4).
Suppose that the velocities of each component are the same at equilibrium \ $({\bf
u}_{1e}\, =\, {\bf u}_{2e})$.  Due to the invariance of the governing equations with respect to the Galilean
transformation we assume, without loss of the generality, that
\ ${\bf u}_{1e}\, =\,{\bf u}_{2e}\, =\,0$.  Suppressing the index  ''$e$''  to avoid double indices,
we get
$$
\cases{\Lambda_{(11)}\ =\ \pmatrix{\rho_1 \,-\, \displaystyle\, {\partial W\over \partial
\mu}}\, \displaystyle\, {I- {\bf ll}^{*} \over \rho^2_1} \, -\, \displaystyle{\partial ^2W\over \partial
\rho^2_1} {\bf ll}^{*} \cr\cr
\displaystyle\Lambda_{(21)}\  = \ \Lambda_{(12)}\  =\ {1\over \rho_1\rho_2}\, {\partial W\over \partial
\mu}(I - {\bf ll}^{*} )  \,- \,\displaystyle\, {\partial ^2W\over \partial\rho_1\partial \rho_2} {\bf
ll}^{*} \cr\cr
\Lambda_{(22)}\  =\ \pmatrix{\rho_2 \ -\ \displaystyle\, {\partial W\over \partial
\mu}}\displaystyle\, {I- {\bf ll}^{*} \over
\rho^2_2} \, -\, {\partial ^2W\over \partial \rho^2_2} {\bf ll}^{*}}
\eqno{(7.12)}
$$
All matrices \ $\Lambda_{(\alpha\beta)}$\  are diagonal.
Now we are able to calculate the dispersion relation (7.11) which is
the determinant of a square matrix of dimension \ $10$. Let us denote
$$
a\  =\ -\, {\partial W\over \partial \mu}, \quad w_{11} \, = \,{\partial ^2W\over \partial
\rho^2_1},\quad w_{12}\  =\ {\partial ^2W\over \partial
\rho_1\partial \rho_2}, \quad w_{22}\  =\ {\partial ^2W\over \partial \rho^2_2}
\eqno{(7.13)}
$$
Suppose that
$$
a\ >\ 0, \quad w_{11}\ >\ 0, \quad w_{22}\ >\ 0, \quad w_{11} w_{22} \,-\, w_{12}^2 \ > \ 0
\eqno{(7.14)}
$$
Taking into account (7.12) - (7.13) we obtain by straightforward calculations of the
determinant (7.11)
$$
\Big(a(\rho_1+\rho_2)\, +\, \rho_1\rho_2\Big)\beta^4_1  \,- \,\Big(a(\rho^2_1w_{11}\, +\, 2\rho_1\rho_2w_{12}
+ \, \rho^2_2w_{22})
\, +\, \rho_1\rho_2 \, (\rho_2w_{22}\, +\, \rho_1w_{11}) \Big) \
$$
$$
 \times (\beta^2_2\, +\, \beta^2_3\, +\, \beta^2_4) \, \beta^2_1 \, + \,
(\beta^2_2\, +\,\beta^2_3\, +\,\beta^2_4)^2(w_{22}w_{11}\, -\, w_{12}^2)\rho^2_1\rho^2_2\  =\ 0
\eqno{(7.15)}
$$
Let \ ${ \mbox{{\boldmath $\nu$}}}\, =\,{\bf l}$,\, ${ \mbox{{\boldmath
$\beta$}}}\, =\,\omega { \mbox{{\boldmath $\nu$}}} \, + \, { \mbox{{\boldmath $\beta$}}}_\sigma$,\,
$\lambda
\,=\,\displaystyle\, {\omega\over \vert{ \mbox{{\boldmath $\beta$}}}_\sigma\vert}$. The dispersion
relation (7.15) takes the form
$$
\Big(a(\rho_1+\rho_2)+\rho_1\rho_2\Big)\lambda^4 \,-\,\Big(a(\rho^2_1w_{11}+2\rho_1\rho_2w_{12} +
\rho^2_2w_{22}) \,+\,\rho_1\rho_2(\rho_2w_{22}+\rho_1w_{11})\Big)\lambda^2
$$
$$+\, (w_{22}w_{11}-w_{12}^2)\rho^2_1\rho^2_2\ =\ 0
\eqno{(7.16)}
$$

\noindent The following result is proved in [21]:

\bigskip

\noindent{\underbar{Theorem 7.1}}
\medskip

\noindent{\it If potential \ $\displaystyle\, W(\rho_1,\rho_2,\mu) $\  satisfies
conditions (7.14), all the roots
\ $\lambda$\  of the polynome (7.16) are real}.

\medskip

\noindent Conditions (7.14) mean that internal energy
$\displaystyle U\, =\, W \,-\, w\, {\partial W\over\partial w}$,  where \ $w = \vert {\bf w} \vert$, is convex
[15-16, 21]. Hence, if
the internal energy is a convex function and the relative velocity \ ${\bf w}$\  is small enough, the
equilibrium state is stable in time direction in the time-space.

It is interesting to note that the Lagrangian of a two-fluid bubbly liquid with incompressible liquid phase
(heterogeneous case) has the same form as (7.1).
Indeed, in [18,19] the potential \ $W$\  for a bubbly liquid is proposed in the form
$$
W\  =\ \rho_2\varepsilon_{20}(\rho_{20}) \, -\, {\rho_{10}\over 2}\,\,m(c)\, \vert {\bf
w}\vert^2
\eqno{(7.17)}
$$
where \ $\rho_2\, =\, c\, \rho_{20}$, $c$ \  is the volume concentration of gas phase and the
index "$0$" means the real density of gas and liquid phase.
Concentration \ $c$\  is expressed by \ $c\, =\, \displaystyle {4\over 3}\, \pi\, R^3\, N$\
where \ $R$\  is the average radius of the bubbles,   $N$\  is the number of bubbles per
unit volume of the mixture. The internal energy per unit mass
\ $\varepsilon_{20}$\  of the gas phase and the virtual mass coefficient \ $m$\  are known functions of \
$\rho_{20}$\  and \ $c$. By introducing the average density of the liquid phase
$$
\rho_1\ =\ \rho_{10}\, (1\, -\, c) \quad \mbox{ with } \quad \rho_{10}\ =\ {\rm const}
$$
we can rewrite the potential (7.17) as
$$
W\ =\ \rho_2\varepsilon_{20} \bigg( {\rho_2\rho_{10}\over \rho_{10}
\,-\,\rho_1}\bigg) \,-\,{\rho_{10}\over 2}\, m\bigg( {\rho_{10} \,-\,\rho_1\over
\rho_{10}} \bigg)\,\vert {\bf w}\vert ^2
$$
The increase of the volume fraction \ $c$\  producing the
interaction between gas bubbles changes not only the coefficient \ $m(c)$\  but also the interfacial energy
\ $\varepsilon_{in}(c)$ \ so that potential \ $W$\  can be generalized to the form
$$
W \ =\ \rho_2\varepsilon_{20}(\rho_{20}) \, -\, {\rho_{10}\over 2}\, m(c) \vert {\bf
w}\vert^2 \, + \,\varepsilon_{in}(c)
\eqno{(7.18)}
$$
Quantity \ $\varepsilon_{in}(c)$ \ produces an additional pressure term due to the
interfacial effect [22]. Therefore, Lagrangian (7.1) describes not only binary
molecular mixtures but also partial cases of suspensions. Case (7.17) is actually
degenerated: \ $w_{11}w_{22} \, -\, w_{12}^2 \, \equiv \, 0$. The introduction of the interfacial
term in (7.18) involves the convexity condition \ $w_{11}w_{22} \, -\, w_{12}^2\, >\, 0$\
provided \ $\displaystyle {\partial^2 \varepsilon_{in}Ê\over \partial c^2} \, > \, 0$.

\appendix
\section{APPENDIX}

For any function \ $\buildrel\circ\over{f} (\lambda  \,,  \,{\bf X})$\  (see
definitions (2.9)-(2.10)) and its image
\ $\overline{f} (t\, , \,{\bf X})$\  in the \ $(t \,, \,{\bf X}) $-coordinates, we obtain the following
relations
$$
{\partial\buildrel\circ\over{f}\over \partial
\lambda}\left({\partial g\over \partial \lambda}\right)^{-1}\ =\ {\partial \overline{f}\over \partial t},
 \quad {\partial \buildrel\circ\over{f}\over \partial
 {\bf X}}\,-\, {\partial g\over \partial {\bf X}}\,{\partial \buildrel\circ\over{f}\over \partial
 \lambda}  \left({\partial g\over
\partial \lambda}\right)^{-1} \  =\ {\partial
\overline{f}\over \partial {\bf X}}
\eqno{(A.1)}
$$

\subsection{Variation of the velocity}

The definition (2.6) of the velocity \ ${\bf u}$\  yields
$$
\buildrel\circ\over{ {\bf u}}\  =\ {\partial { \mbox{{\boldmath $\phi$}}}\over
\partial\lambda}\, \left({\partial g\over \partial \lambda}\right)^{-1}
$$
In Lagrangian coordinates,
perturbation of \ ${\bf u}$\  is represented by the formula
$$
\buildrel\sim\over {\bf u}(\lambda
\,, \,{\bf X} \,, \,\varepsilon)\ =\ {\partial {\bf \Phi}\over \partial \lambda} (\lambda \,,
\,{\bf X} \,, \,\varepsilon)\, \left( {\partial G\over \partial \lambda}\,(\lambda \,, \,{\bf X}
\,, \,\varepsilon)\right)^{-1}
$$
Taking the derivative of \ $\buildrel\sim\over{\bf u}$\  with respect to
\ $\varepsilon$\  at \ $\varepsilon\, = \,0$\  and using (A.1), we get
$$
\buildrel\sim\over{\delta} {\bf u}\  =\ {\partial \buildrel\circ\over{ \mbox{{\boldmath
$\xi$}}}\over  {\partial \lambda}}\, \left( {\partial g\over \partial \lambda}\right)^{-1}
\,-\, \buildrel\circ\over{\bf u}\,{\partial \buildrel\circ\over\tau\over\partial \lambda}\left({\partial
g\over \partial \lambda}\right)^{-1}\  =\ {\partial \overline{ \mbox{{\boldmath $\xi$}}}\over  {\partial t}}
\,-\, \overline {\bf u}\,{\partial \overline {\tau} \over \partial t} \ =\ {d { \mbox{{\boldmath
$\xi$}}}\over {d t}} \,-\, {\bf u}\, {d \tau\over d t}
$$
where
\ $\displaystyle\, {d\over dt}\, =\, {\partial \over \partial t} \, +\, {\bf
u}^{*}\, \bigtriangledown^{*} $\   is the
material derivative.
 We obtain then
$$
\buildrel\sim\over\delta {\bf v}\  =\ (I \, -\, {\bf v} {\bf l}^{*} ) \,{\partial { \mbox{{\boldmath
$\zeta$}}}\over
\partial {\bf z}}\, {\bf v}
\eqno{(A.2)}
$$
where \ $\displaystyle {\bf v}^{{*}}\,  = \, (1, {\bf u }^{{*}})$, $I$ \  is the identity tensor and
\  ${\bf l}^{*} \, =\, (1,0,0,0)$.

\subsection{Variation of the deformation gradient}

In Lagrangian
coordinates \ $(\lambda, {\bf X})$, the perturbation of the deformation gradient is given by (2.6)
$$
\buildrel\sim\over F\, (\lambda\,, \,{\bf X} \,, \,\varepsilon)\ =\ { \partial
{\bf \Phi}\over
\partial {\bf X}} \,-\,{\partial {\bf \Phi}\over
\partial \lambda} {\partial G\over \partial {\bf X}}\, \left( \,{\partial G\over \partial
\lambda}\, \right)^{-1}
$$
Combination of the derivative of \ $\buildrel\sim\over F$\  with
respect to \ $\varepsilon$\  at \ $\varepsilon\, =\,0$\  and relation (A.1) gives
$$
\buildrel\sim\over{\delta} F\ =\ {\partial \buildrel\circ\over{ \mbox{{\boldmath
$\xi$}}}\over \partial {\bf X}}
\,-\,{\partial \buildrel\circ\over{ \mbox{{\boldmath $\xi$}}}\over \partial \lambda}\, {\partial g\over
\partial {\bf X}}\, \left({\partial g\over\partial\lambda} \right)^{-1} \,-\, {\partial {\bf \Phi}\over
\partial \lambda}\, {\partial
\buildrel\circ\over{\tau}\over \partial {\bf X} }\, \left({\partial g\over \partial {\lambda}} \right)^{-1}
\, +\, {\partial { \mbox{{\boldmath $\phi$}}}\over \partial \lambda}\, {\partial g\over \partial {\bf
X}}\, \left({\partial g\over
\partial \lambda}\right)^{-2} \,{\partial
\buildrel\circ\over{\tau}\over \partial \lambda }\
$$
$$
 =\  {\partial \overline{ \mbox{{\boldmath $\xi$}}}\over
\partial {\bf X}}\,-\, \overline{\bf u}\, {\partial \overline{\tau}\over \partial {\bf X}}
\  = \ \bigg(\, {\partial { \mbox{{\boldmath $\xi$}}}\over \partial
{\bf x}} \, -\, {\bf u}\, {\partial {\tau}\over \partial {\bf x}}\bigg)\, F
\eqno{(A.3)}
$$

\subsection{Variation of the density}

The mass conservation is represented by the formula
$$
\rho\, \det\, F\  =\ \rho_0({\bf X})
$$
The perturbation of \ $\rho$\  is in the form
$$
\buildrel\sim\over{\rho} (\lambda \,, \,{\bf X}\,,
\,\varepsilon)\, \det\, \buildrel\sim\over{F} (\lambda \,, \,{\bf X} \,, \,\varepsilon)\
=\ \rho_0( {\bf X)}
$$
and consequently
$$
\buildrel\sim\over{\delta}\rho\, \det \,\buildrel\circ\over F
\, +\, \buildrel\circ\over\rho\, \buildrel\sim\over\delta (\det\,\,F)\  =\ 0
\eqno{(A.4)}
$$
By using the Euler-Jacobi identity,
$$
\delta\, (\det\, F)\,  =\, \det \, F\,tr\, (F^{-1}\buildrel\sim\over\delta\, F)
$$
relations (A.3) and (A.4) lead to
$$
\buildrel\sim\over\delta \rho\ = \ -\, \,tr\, \bigg(\rho( I  \,- \,{\bf v} {\bf l}^{*})
\, {\partial { \mbox{{\boldmath $\zeta$}}}\over \partial{\bf z}} \bigg)
\eqno(A.5)
$$

\subsection{Variation of specific quantities}

Along each trajectory the specific quantities $a_k$ are constant. Consequently,
\ $a_k\, =\, a_{0k}({\bf X})$. The perturbation of \ $a_k$\  is such that
\ $\buildrel\sim\over a_k \, (\lambda \,, \,{\bf X}\,, \,\varepsilon)\,
=\, a_{0k}( {\bf X})$. Hence,
$$
\buildrel\sim\over\delta a_k \  = \ 0
\eqno(A.6)
$$

\section{APPENDIX}

Formulae (A.2) and (A.5) yield directly the variation of
\ $\displaystyle\, {\bf j}_0$
$$
\buildrel\sim\over\delta {\bf j}_0 \ = \ \left(tr\, \left(\left( {\bf v}{\bf l}^{*} \,-\,I
\right) \,{\partial  { \mbox{{\boldmath $\zeta$}}}\over \partial {\bf z}}\right) \,I \,-
\, ( {\bf v l}^{*}
\,- \,I ) {\partial { \mbox{{\boldmath $\zeta$}}}\over
\partial {\bf z}} \right) \,{\bf j}_0
$$
$$  =\ \left({\partial { \mbox{{\boldmath $\zeta$}}}\over \partial {\bf z}} \,
-\, ( Div\,\,{ \mbox{{\boldmath $\zeta$}} }) \,I \right)\, {\bf j}_0
\eqno{(B.1)}
$$
Let us consider the variations
\ $\buildrel\sim\over\delta {\bf j}_k$,\, $k\, =\,1,\dots,m$\
with \ ${\bf j}_k\, =\, a_k {\bf j}_0$\  where the scalar fields \ $a_k$\  are such that \
$a_k\, =\, a_{k 0}( {\bf X})$.
Due to (A.6)
 \ $\buildrel\sim\over \delta a_k\, =\, 0$\  and \ $\buildrel\sim\over \delta {\bf j}_k\, = \,a_k
\buildrel\sim\over \delta {\bf j}_0$.
Consequently
$$
\buildrel\sim\over \delta\, {\bf j}_k\  =\ \bigg( {\partial { \mbox{{\boldmath $\zeta$}}}\over \partial
{\bf z}} \, -\, ( {\rm Div} \,{ \mbox{{\boldmath $\zeta$}}} ) \,I \bigg)\, {\bf j}_{k}, \quad
k \ =\ 0,\dots,m
\eqno{(B.2)}
$$
Hence, (2.12) involves
$$
\buildrel\wedge\over{\delta}{\bf j}_k\  = \ \bigg({\partial { \mbox{{\boldmath $\zeta$}}}\over
\partial {\bf z}} \, -\, ( {\rm Div} \,{ \mbox{{\boldmath $\zeta$}}})\, I \bigg)\, {\bf j}_k
\, -\, {\partial {\bf j}_k\over \partial {\bf z}}\, { \mbox{{\boldmath $\zeta$}}}, \quad
k\  =\ 0,\dots,m
\eqno{(B.3)}
$$

\section{APPENDIX}

Let \ ${ A  \,,  \,B}$\  be two
linear transformations
depending on \ ${\bf z}$. Let us define the linear form \ $\displaystyle tr\, \bigg( { A}\, {\partial {
B}\over \partial {\bf z}}\bigg)$\  such that for any
constant vector field \ ${\bf a}$,
\ $\displaystyle tr\, \bigg( { A}\, {\partial { B}\over \partial {\bf z}}\bigg) \,{\bf
a}\, \equiv\, tr\, \bigg( { A}\, {\partial { B}{\bf a}\over \partial {\bf z}}\bigg)$.
As a consequence, we get
$$
{\rm Div} \,( { A\,\,B})\  =\ {\rm Div} \,( { A} ) \, { B} \,+ \,tr \,\bigg( { A}\, {\partial {B}\over
\partial {\bf z}}\bigg)
\eqno{(C.1)}
$$
Indeed,
$$
{\rm Div} \, ( { A}\, { B}  \,{\bf a})\  =\ {\rm Div} \,( { A}) \,{ B} \,{\bf a} \,+
 \,tr \, \bigg( { A} \,{\partial { B}{\bf a}\over
\partial {\bf z}}\bigg) \
=\ \left( ({\rm Div} \, { A}) \, { B} + \,tr \,\left( { A} \,{\partial { B}\over
\partial {\bf z}} \right) \right) {\bf a}
$$
Let us also recall the following useful formulae
for any vector fields \ ${\bf b}( {\bf z})$,\, ${\bf c}( {\bf z})$
$$
{\rm Div} \,( {\bf b}\, {\bf c}^{*} )\  =\ {\bf c}^{*} \,{\rm Div} \,{\bf b  \,+\, b}^{*}
\bigg(\, {\partial {\bf c}\over \partial {\bf z}}\bigg)^{*}
\eqno{(C.2)}
$$
$$
{\rm Div} \,( {\bf b^{*} \, c} \,I)\  =\ {\partial \over \partial {\bf z}}\, ( {\bf b^{*}
\, c})\  =\ {\bf b^{*} }\, {\partial {\bf c}\over \partial {\bf z}} \, + \,{\bf c^{*} } \,{\partial
{\bf b}\over \partial {\bf z}}
\eqno{(C.3)}
$$
By using formulae (C.1) - (C.3), we have the following identities
$$
{\rm Div}\,( {\bf K}^{*} _k {\bf j}_k\,\,I)\ \equiv\ {\bf K}^{*}
_k\, {\partial {\bf j}_k\over \partial {\bf z}} \, +\, {\bf j}_k^{*} {\partial {\bf K}_k\over \partial {\bf
z}}
\eqno{(C.4)}
$$
$$
{\rm Div}\, ( {\bf j}_k {\bf K}_k^{*})\ \equiv\ - \, {\bf
K}^{*}_k\,{\rm Div}\,{\bf j}_k \, -\, {\bf j}_k^{*}
\bigg({\partial {\bf K}_k\over \partial {\bf z}}\bigg)^{*}
\eqno{(C.5)}
$$
$$
{\rm Div}\, \bigg(A^{*}_k\, {\partial {\bf j}_k\over \partial {\bf
z}}
\bigg)\ \equiv\ {\rm Div}\, (A^{*}_k)\, {\partial {\bf j}_k\over \partial {\bf z}}
\, +\,  tr\, \Bigg(A^{*}_k\, {\partial \over \partial {\bf z}} \bigg({\partial {\bf j}_k\over \partial {\bf
z}}\bigg)\Bigg)
\eqno{(C.6)}
$$
$$
{\rm Div}\, (\Lambda \, I)\  \equiv\ - \sum^m_{k=0}\,\,\left( {\partial \Lambda\over
\partial {\bf j}_k}\, {\partial {\bf j}_k\over \partial {\bf z}} \, +\, tr \left(\, {\partial \Lambda\over
\partial \, \left(\displaystyle{\partial {\bf j}_k\over \partial {\bf z}}\right)\, }\, {\partial \over \partial
{\bf z}}\bigg( {\partial {\bf j}_k\over \partial {\bf z}}\bigg)\right)\right) \, -\, {\partial
\Lambda\over \partial {\bf z}}
$$
$$
\equiv\ - \, \sum^m_{k=0}\, \left(  \,{\bf K}^{*} _k
{\partial {\bf j}_k\over \partial {\bf z}} \, +\, tr \left(\,\,A^{*}_k{\partial \over \partial {\bf
z}}\left( {\partial {\bf j}_k\over \partial {\bf z}}\right)\right)\right) \, - \, {\partial \Lambda\over
\partial {\bf z}}
\eqno{(C.7)}
$$
By adding (C.4) - (C.7) we obtain
$$
{\rm Div} \, \left( \, \sum^m_{k=0}\, \left( {\bf K}_k^{*} {\bf j}_k\, I \,-\, {\bf j}_k
{\bf K}_k^{*}  \,+\, A^{*}_k\, {\partial {\bf j}_k\over \partial {\bf z}}\right) \,-\, \Lambda
\, I\right) \,+\, {\partial \Lambda\over \partial {\bf z}}
$$
$$
+\, \sum^m_{k=0}\,\,{\bf K}_k^{*} \,{\rm Div} \, {\bf j}_k\, -\, \sum^m_{k=0}\,\,{\bf j}_k^{*}
\Bigg(\, {\partial {\bf K}_k\over \partial {\bf z}} \, - \,\bigg( {\partial {\bf K}_k\over \partial {\bf
z}}\bigg)^{*} \Bigg)\ \equiv\ 0
$$
which proves Theorem 4.1.

\end{document}